\documentclass{svproc}

\usepackage{url}

\usepackage{verbatim}
\usepackage{graphicx}
\usepackage{lineno}
\usepackage{color}
\usepackage{caption}
\usepackage{subfig}
\usepackage{subfig}
\usepackage{adjustbox}
\usepackage{tabularx}
\usepackage{csquotes}
\usepackage{cite}
\usepackage[T1]{fontenc}
\usepackage{hyperref}
\usepackage{textcomp}

\begin{document}
\mainmatter              
\title{ Assembly, testing, and installation of mPMT photosensor for the Water Cherenkov Test Experiment}
\titlerunning{mPMT for WCTE}  
%

\author{
	M. Gola\thanks{Corresponding Author: mgola@triumf.ca}\inst{1}, 
	M. Barbi\inst{4}, V. Berardi\inst{8}, A. Buchowicz\inst{5}, N. Buril\inst{3}, 
	L. Cook\inst{1}, S. Cuen-Rochin\inst{9}, 
	G. DeRosa\inst{8}, , P. de Perio\inst{11}, K. Dygnarowicz\inst{5}, 
	B. Ferrazzi\inst{4}, A. Fiorentini\inst{3}, 
	C. S. Garde\inst{10}, G. Galiński\inst{5}, K. Graham\inst{3}, 
	R. Gornea\inst{3}, M. Hartz\inst{1}, 
	J. Holeczek\inst{6}, S. Jagtap\inst{10}, 
	M. Kala\inst{3}, D. Karlen\inst{2}, 
	S. Kothekar\inst{10}, L. Koerich\inst{4}, 
	N. Kolev\inst{4}, A. Konaka\inst{1}, 
	A. Kulkarni\inst{10}, J. Kowalewski\inst{5}, 
	R. Kurjata\inst{5}, X. Li\inst{1}, 
	T. Lindner\inst{1}, P. Lu\inst{1}, 
	A. Mache\inst{10}, J. Marzec\inst{5}, 
	I. Nikonov\inst{1}, M. Nurek\inst{5}, 
	W. Obrębski\inst{5}, S. Patil\inst{10}, 
	G. Pastuszak\inst{5}, B. Piotrowski\inst{5}, 
	J. Rimmer\inst{1}, B. Roskovec\inst{7}, 
	A. C. Ruggeri\inst{8}, A. Rychter\inst{5}, 
	K. Satao\inst{10}, N. Sharma\inst{10}, 
	A. Stockton\inst{1}, S. Yousefnejad\inst{4}, 
	T. Yu\inst{1}, M. Ziembicki\inst{5}
}

\authorrunning{Mohit Gola} 
%
%
\institute{
	$^{1}$TRIUMF, Vancouver, Canada\\
	$^{2}$University of Victoria, Victoria, Canada\\
	$^{3}$Carleton University, Ottawa, Canada\\
	$^{4}$University of Regina, Regina, Canada\\
	$^{5}$Warsaw University of Technology, Warszawa, Poland\\
	$^{6}$University of Silesia, Katowice, Poland\\
	$^{7}$Charles University, Prague, Czechia\\
	$^{8}$INFN - Sezione di Bari, Bari, Italy\\
	$^{9}$Tecnologico de Monterrey, Culiacan, Mexico\\
	$^{10}$ Vishwakarma Institute of Information Technology, Pune, India\\
	$^{11}$ Kavli IPMU (WPI), UTIAS, The University of Tokyo, Kashiwa, Japan}	

\maketitle              

\begin{abstract}

The multi-Photomultiplier Tube (mPMT) photosensors will be used in the Water Cherenkov Test Experiment (WCTE) to efficiently detect the photons produced in the whole detector. One of the aims behind the development of WCTE is to test the technology and implement it in future water Cherenkov experiments such as the Hyper-Kamiokande experiment and its Intermediate Water Cherenkov Detector. Each mPMT is built using nineteen 3-$\rm inch$ PMTs arranged on a semi-spherical support matrix. In this paper, we describe the design and manufacture of the mechanical components, the procedures for casting an optical gel between PMTs and acrylic cover, and the overall assembly procedure of the mPMTs. Details of the electronics used in the mPMT modules are not included in this paper and will be presented in a separate publication. We also report on the R\&D performed on the selection of the optical gel ratio along with transmittance measurements and the reflectance measurements performed on the aluminium reflector. We also present the optical tests performed on the mPMT module using a 405~$\rm nm$ LED and the resulting increase in the effective photosensitive area by surrounding the PMTs with a reflector. A summary of the production and installation of the mPMTs for the WCTE is also presented in this paper.
\keywords{mPMTs,  Water Cherenkov, WCTE, Hyper-Kamiokande}
\end{abstract}


\section{Introduction}
The Hyper-Kamiokande (Hyper-K) experiment \cite{HK1, HK2, HK3} is based on the successful long baseline neutrino project in Japan and a next-generation experiment to Super-Kamiokande (Super-K)  \cite{SK1, SK2, SK3}. The Hyper-K far detector will utilize a 258-kiloton water tank located in the largest man-made experimental cavern. One of the main objectives of Hyper-K is the precise investigation of the leptonic CP violation using accelerator neutrinos \cite{CP-Violation1, CP-Violation2}. The experiment will also search for proton decay and study solar neutrino oscillations among others \cite{protonDecay}. For the precise measurement of the accelerator neutrino oscillation, it is crucial to measure the neutrino flux and its interaction cross-section at the point where the neutrino beam is generated before the effects of neutrino oscillation become significant. For this purpose, the Intermediate Water Cherenkov Detector (IWCD) has been proposed as a new near detector, consisting of a water tank with a capacity of 1 kiloton \cite{IWCD}.  The Hyper-K far detector and IWCD will use multi-photomultiplier tubes (mPMTs) as one of the main photon detection systems \cite{FD-mPMT1, FD-mPMT2, FD-mPMT3, FD-mPMT4, FD-PMT, FD-mPMT-Electronics}. The design of the Hyper-K and IWCD mPMTs is inspired by the optical modules developed for the KM3NeT experiment \cite{KM3Net1, KM3Net2}. The number of mPMTs used in Hyper-K and IWCD is respectively 800 and 400. The Hyper-K far detector mPMTs are slightly different from IWCD mPMTs and are not part of this paper.   

\begin{figure}
\includegraphics[scale=0.25]{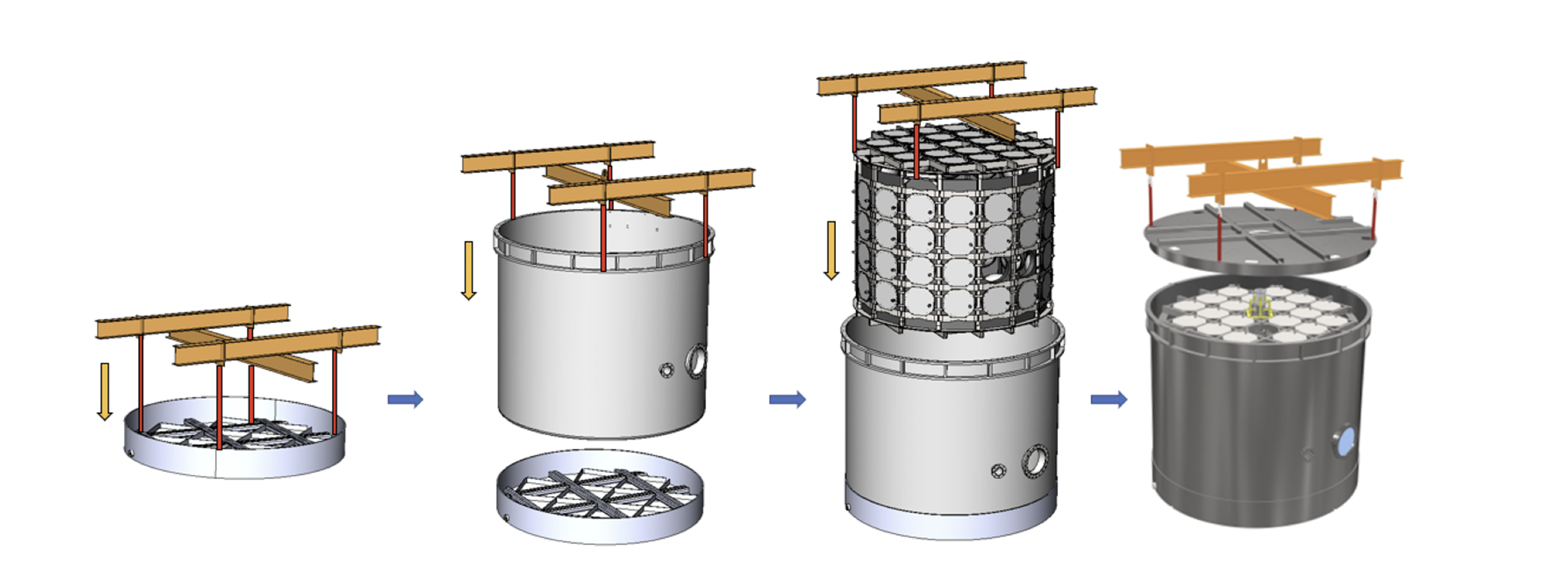}
\caption{Installation sequence of the WCTE detector in the experimental area at CERN. Approximately 100 mPMTs are pre-mounted onto a cylindrical steel support structure, which is then lowered into the water tank as shown in the third panel. A close-up view of the mounted mPMTs is provided in Fig.~\ref{fig:WCTE_Structure}.}
\label{fig:WCTE_Detector}
\end{figure}

The construction of the Water Cherenkov Test Experiment (WCTE) with 40 tons of water tank capacity started in June 2024 at CERN \cite{WCTE0, WCTE1, WCTE2}. One of its purposes is to comprehensively understand the technology and confirm the efficient operation and performance of the IWCD. The water tank for WCTE will adopt a cylindrical shape, with dimensions approximately 4 meters in height and diameter, facilitating the propagation of particles for physics investigations. The design and installation procedures of the WCTE detector are illustrated in Figure~\,\ref{fig:WCTE_Detector}. Situated in the East Area T9 beamline at the European Centre for Nuclear Research (CERN), the WCTE experiment will focus on measuring water Cherenkov radiation emitted by low momentum (200-1000 $\rm MeV/c$) $\pi^\pm$, $\mu^\pm$, $e^\pm$, and $p^+$ particles produced in the secondary beam at the T9 area. WCTE will contain 97 mPMTs, including four Hyper-K style mPMTs, and each mPMT unit encompasses nineteen 3-inch PMTs enclosed within a water-resistant vessel alongside the necessary readout electronics. \\
The structure of the paper is organized as follows: Section~\ref{s:mPMT Design}  provides an overview of the mPMT photosensor design, detailing the mechanical components involved in the assembly. The step-by-step assembly procedure of the mPMT is outlined in Section~\ref{s:mPMT Assembly Procedure}. Section~\ref{s:MTS} discusses the relative efficiency measurements performed at the mPMT testing facility. Section~\ref{s:Production_Summary} summarizes the production and installation of the 100 mPMTs constructed for WCTE. Finally, the conclusions are presented in Section~\ref{s:Conclusions}. This paper does not cover the design, fabrication and testing of the electronics within the mPMT module.   This information will be covered in a future paper.  For this paper, we simply note that the mPMT electronics generate the high voltage (HV) to operate the PMTs and use fast digitization to precisely measure photo-electron arrival times and charges.



\section{mPMT Photosensor Design}\label {s:mPMT Design}

\begin{figure}
    \centering
    \begin{minipage}{0.45\textwidth}
        \centering
        \includegraphics[width=\linewidth]{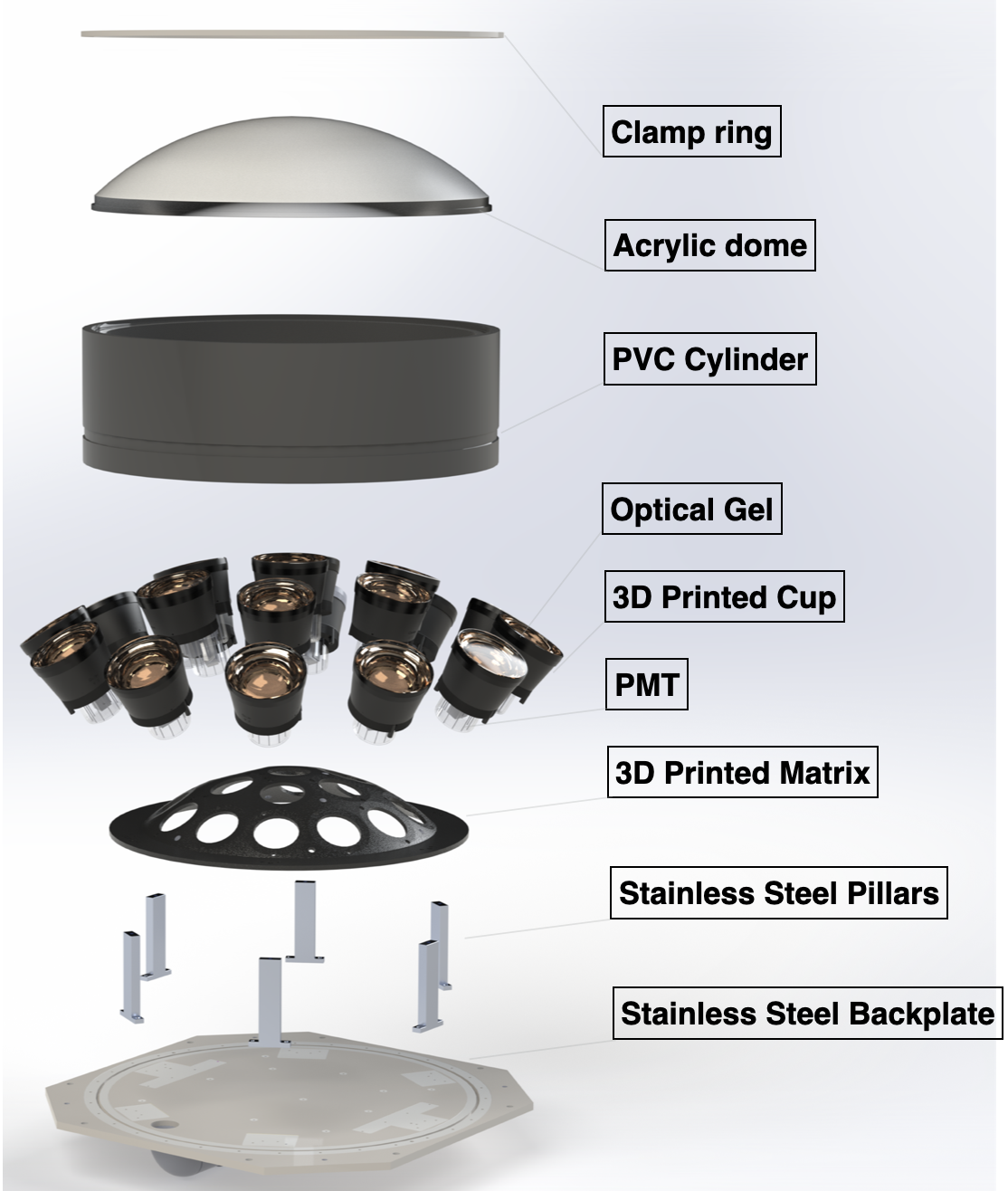}
        \caption*{(a) Ex-situ style mPMT}
    \end{minipage}
    \hfill
    \begin{minipage}{0.44\textwidth}
        \centering
        \includegraphics[width=\linewidth]{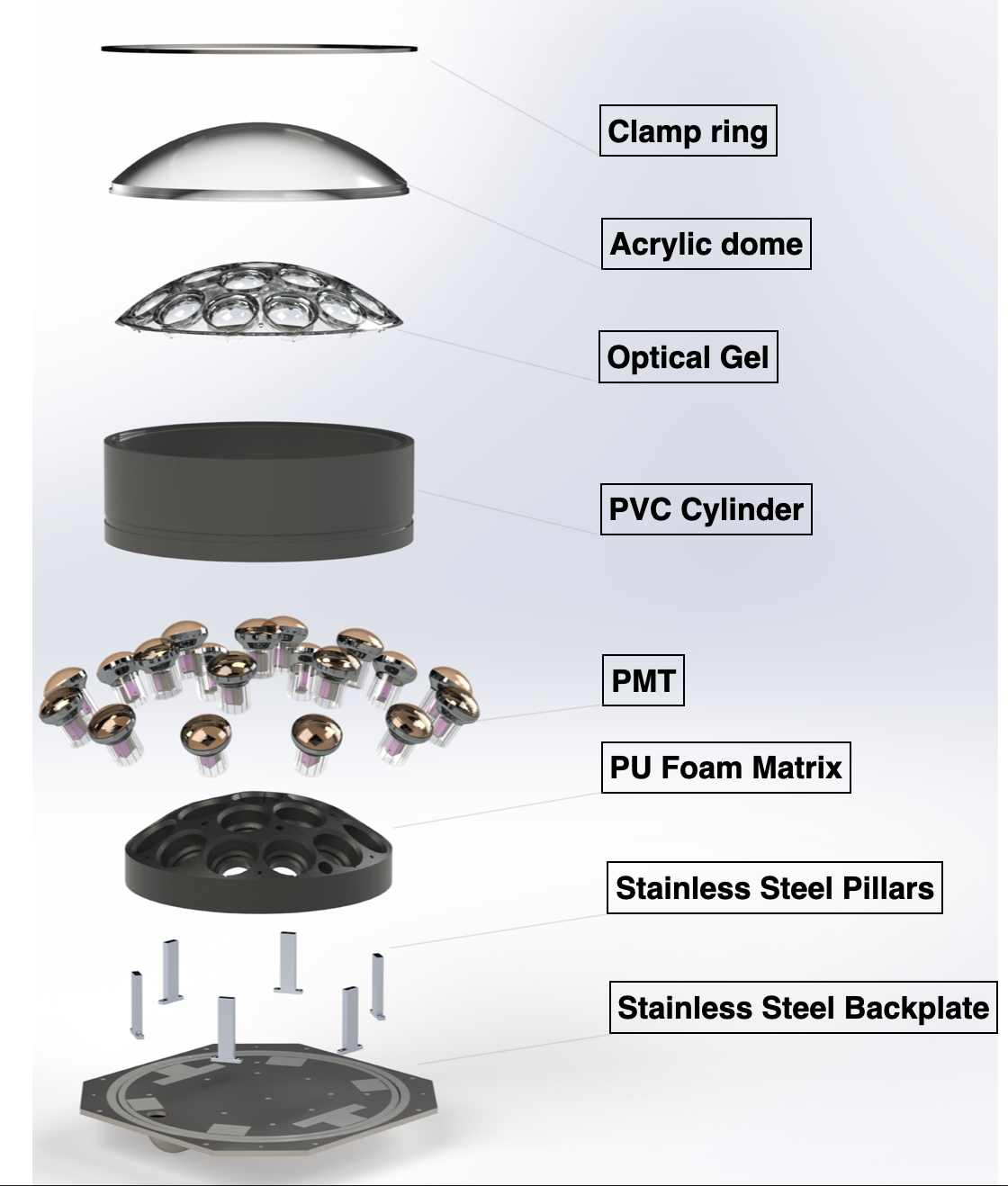}
        \caption*{(b) In-situ style mPMT}
    \end{minipage}
    \caption{The design of the IWCD mPMT.}
    \label{fig:WCTE_mPMT}
\end{figure}

The majority of the mPMTs for the WCTE follow the design of the IWCD mPMTs. Therefore, these mPMTs will be referred to as IWCD-style mPMTs~\cite{IWCD-mPMT} throughout the paper, and this terminology will be consistently used. The design of the mPMTs is such that the PMTs are installed on the support matrix, which sits on a stainless steel backplate, surrounded by a PVC cylinder, and covered with an acrylic dome. There is an optical gel between the PMTs and acrylic dome to ensure similar indices of refraction along the path of photons and reduce reflections at different boundaries. Two assembly strategies have been developed to ensure optical gel contact with the acrylic dome of the mPMT vessel: \enquote{ex-situ} and \enquote{in-situ}. In the ex-situ gelling approach, the gel is cast onto individual mPMTs using a mould before being installed into the complete mPMT. Conversely, in the in-situ gelling method, PMTs are first positioned within the mPMT vessel, followed by gel pouring between the PMTs and the acrylic dome.  The structure of the current ex-situ and in-situ mPMT photosensors are shown in Figure~\ref{fig:WCTE_mPMT} (a) \& (b). 
The mechanical components to assemble the mPMT photosensor are a stainless steel backplate, a clamp ring, six pillars, a 3D-printed/PU-foam PMT support matrix, a PVC cylinder, an acrylic dome, and a mainboard. Each mPMT module is equipped with six continuous-output LEDs for photogrammetry targeting and three fast-pulsing LEDs for in-situ detector calibration. These LEDs are connected to light pipes from the mainboard, directing light to the front of the detector. The LED light sources are essential for time calibration and the study of scattering and reflections within the detector and also serve as beacons for the photogrammetry system.


The IWCD-style mPMT mechanical design is driven by various factors such as the WCTE and IWCD detector design, physics requirements, and cost considerations, some of which are discussed herein. The mPMT mechanical design shall conform to certain requirements; 
 \begin{itemize}
\item The mPMT vessel should be water-tight up to a depth of 4~$\rm m$ (10~$\rm m$) for WCTE (IWCD).\\
\item The mPMT vessel material should not seriously degrade the WCTE/IWCD water quality (both ultra-pure water and $\sim$0.1wt\% Gd-doped  \cite{SK-Gd} ultra-pure water will be used in WCTE). \\
\item The complexity of the design should be minimal to ease the manufacturing and mPMT assembly process.
 \end{itemize}
The design of the reflector ring, 3D printed/PU-foam support matrix, PVC cylinder, and acrylic dome are described in the following subsections. 
 
\begin{figure}
\begin{center} \includegraphics[scale=0.2]{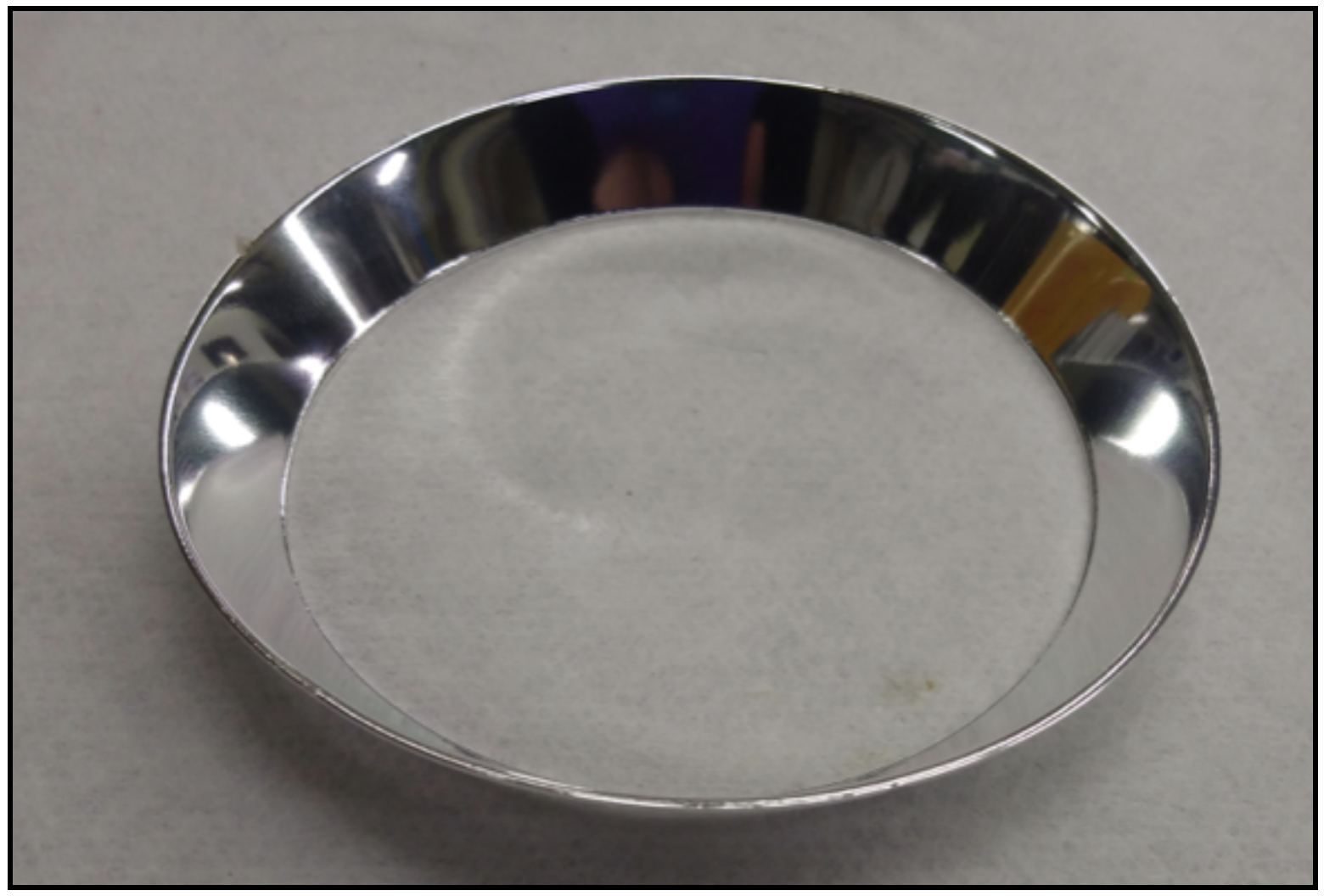} \end{center}
\vspace{-10pt}
\caption{Aluminum reflectors from the company Nata.}
\label{fig:Reflector_Step}
\end{figure}

\vspace{-10pt}
\subsection{Aluminium Reflector}
A small conical piece of reflective metal is placed around each PMT to increase light yield. The Nata Lightning Company commercially manufactures these reflectors with the design prepared by us, as shown in Figure~\,\ref{fig:Reflector_Step}. Section \ref{s:Efficiency_Measurement_w/wo_Reflector} compares the photosensitive area between PMTs with and without reflectors. To determine the most suitable material based on reflectance, an absolute specular reflectance measurement was carried out using the PerkinElmer Lambda 950 spectrometer. Five different samples were analyzed across the wavelength range of 200~$\rm nm$ to 500~$\rm nm$. The samples tested included Analod MIRO IV reflector sheets, Anolux MIRO IV reflector sheets, a reflector sample provided by INFN made from polished aluminium, Anodized WF (without sealing), and Anodized FB (with sealing) from Nata company. 

\begin{figure}
\begin{center} \includegraphics[scale=0.35]{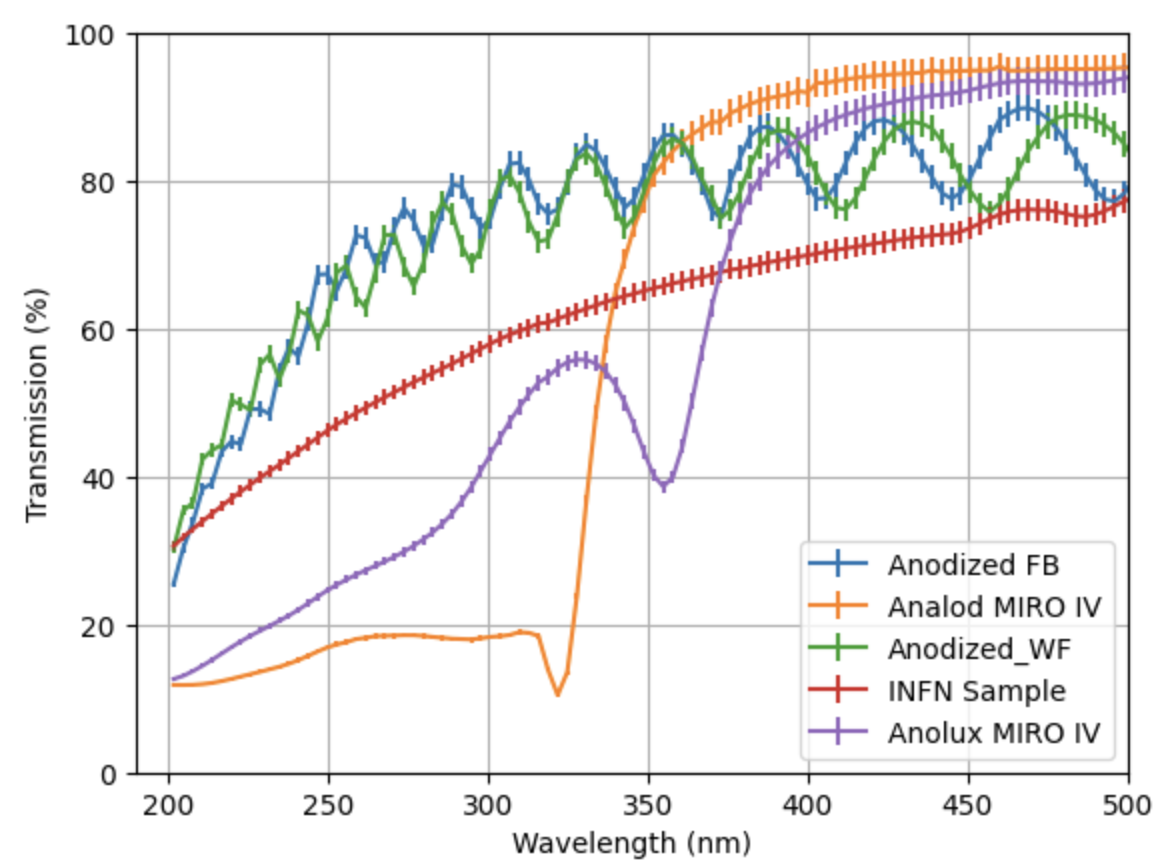} \end{center}
\vspace{-10pt}
\caption{Reflectance measurement on different samples.}
\label{fig:Reflector_measurement}
\end{figure}

Data analysis indicates that anodized aluminium exhibits higher reflectance as shown in Figure~\,\ref{fig:Reflector_measurement}, compared to the other samples in the 200~$\rm nm$ to 500~$\rm nm$. Beyond 350~$\rm nm$, its reflectance is only slightly lower than that of the Anolux and Analod MIRO IV samples. Moreover, anodized aluminium is significantly more cost-effective than MIRO IV material for reflector applications. Therefore, we opted to use anodized aluminum, which provides similar reflectance to other materials but is more cost-effective and efficient to produce.

\subsection{PMTs Support Matrix} \label {s:matrix}

The PMTs are placed on a support matrix that holds them at a specific angle and distance relative to the dome inside the mPMT. The position and angle of the PMTs help to ensure good optical contact between the PMTs with the optical gel on top and the acrylic dome cover.


\begin{figure}
    \centering
    \subfloat[Ex-situ matrix]{%
        \includegraphics[scale=0.246]{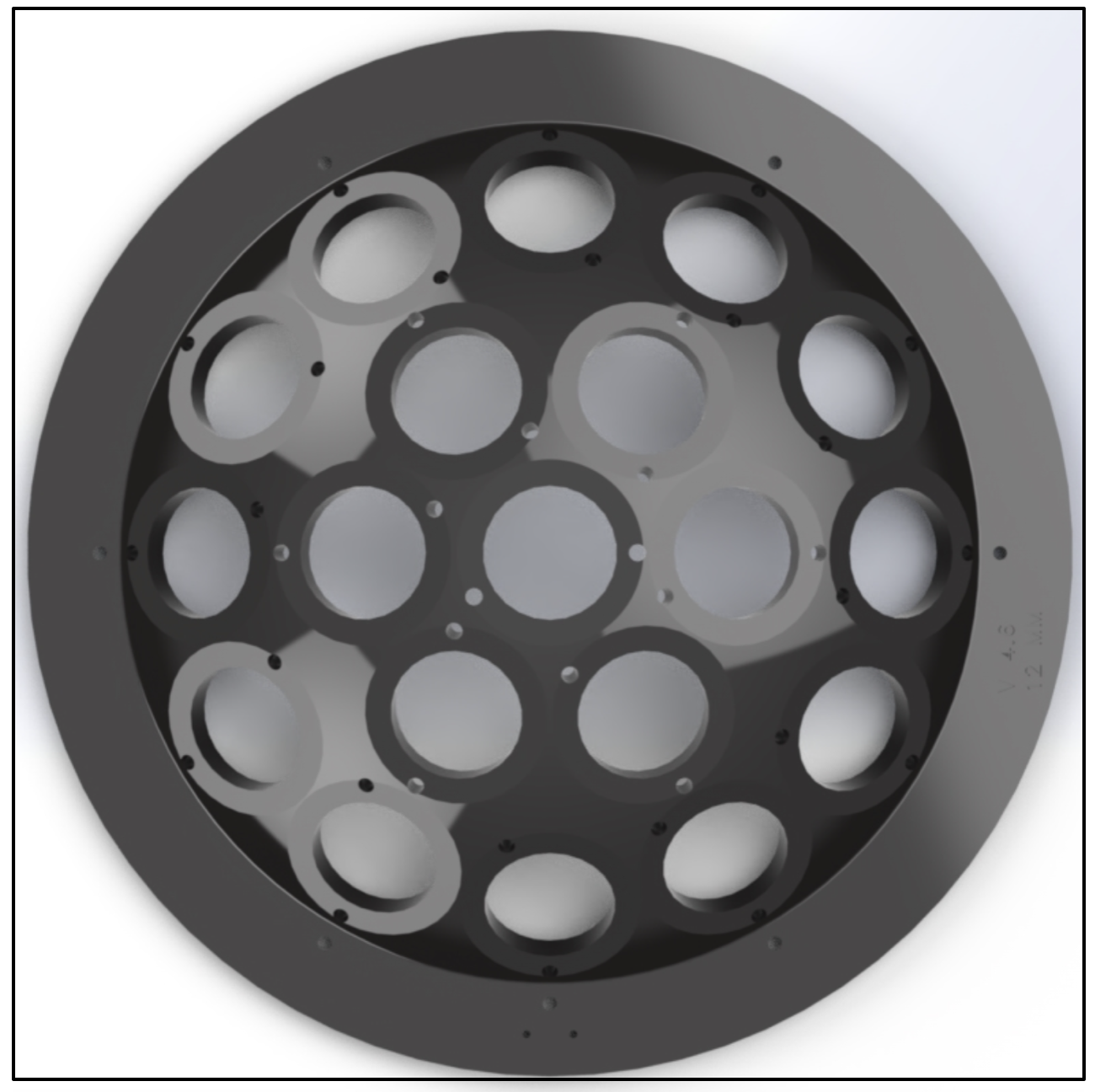}%
        \label{fig:exsitu_matrix}
    }
    \hfill
    \subfloat[In-situ matrix]{%
        \includegraphics[scale=0.25]{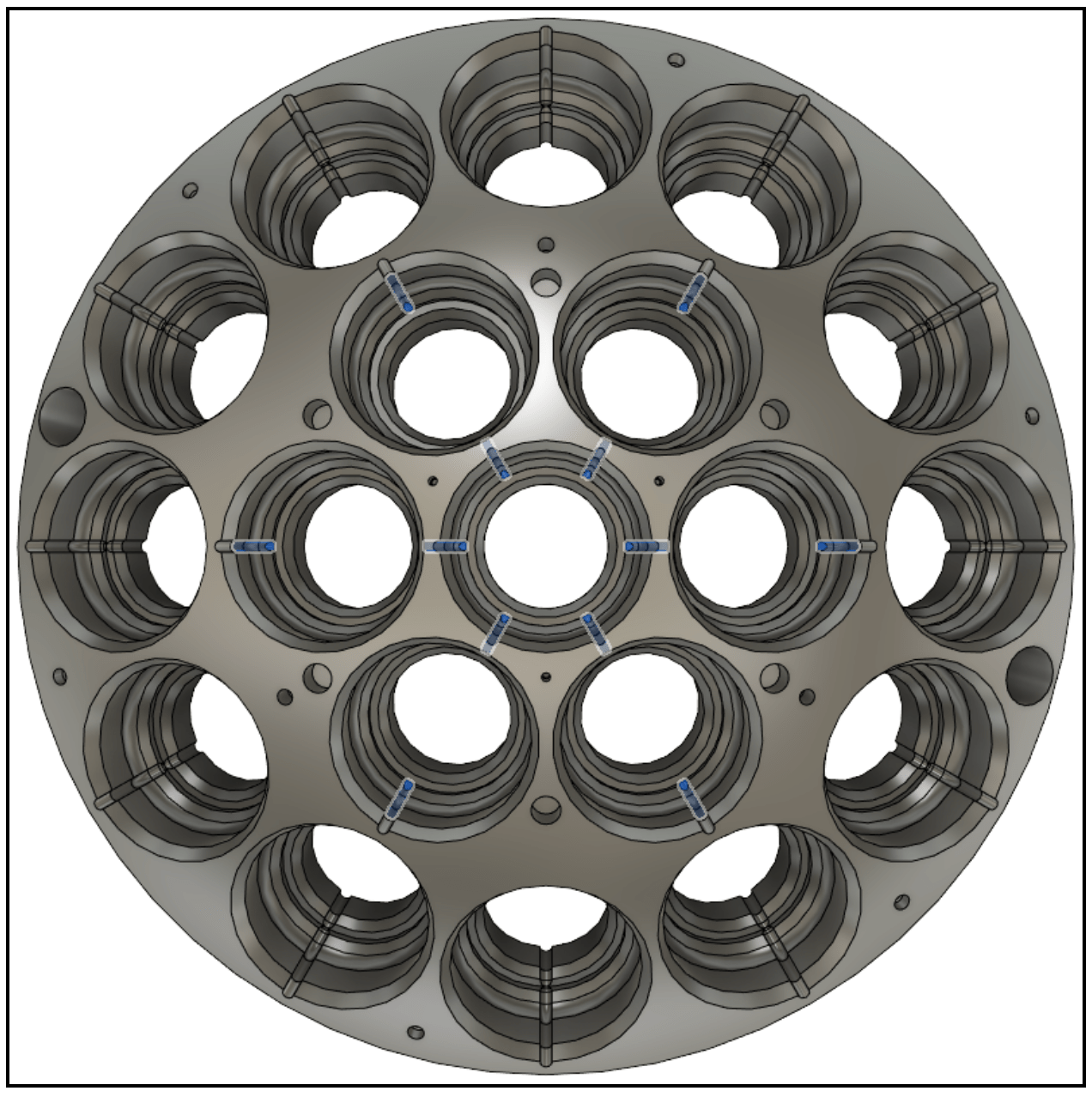}%
        \label{fig:insitu_matrix}
    }
    \caption{The design of the ex-situ (a) and the in-situ (b) matrix for IWCD-style mPMT.}
    \label{fig:Matrix}
\end{figure}

For the ex-situ assembly strategy, the support matrix is produced using a 3D printer with black Acrylonitrile Butadiene Styrene (ABS) material in two pieces which are glued together, as shown in Figure~\,\ref{fig:Matrix}~(a). The ABS material is rigid, and it will not come in contact with ultra-pure water, so compatibility is not an issue. The ex-situ support matrix is 6~$\rm mm$ thick to ensure it can handle the weight of the gelled PMTs and compression of the dome without sagging. For the in-situ assembly strategy, the matrix is made using two-component Polyurethane (PU) foam along with black dye added to the mixture to reduce the surface reflectivity. A cylindrical foam pucked is made using this foam, which is machined using Computer Numerical Control (CNC) machining to make the holes for the PMT seating. The in-situ matrix is shown in Figure~\,\ref{fig:Matrix}~(b). To prepare a PU foam matrix, the process begins with thorough mould preparation. The mould is treated with paraffin wax, which is melted and evenly distributed using a heat gun. An even wax coating is essential to avoid any issues during demolding. The foam mixing process involves aerating both components A and B before measurement to ensure consistency in the chemical reaction. The foam block is created in two stages, with a 40-minute interval between pours. For the first pour, the correct amounts of component A, black dye, and component B are measured and quickly mixed due to the foam’s short pot life. The mixture is blended using a paint shaker and then poured into the mould. The second pour begins 30 minutes after the first and follows the same mixing process. The mould is then left to cure for about four hours. Once cured, the foam shrinks slightly, allowing it to be removed by flipping the mould over. If necessary, gently dropping the mould can help loosen the foam. Then the foam puck is machined using CNC machining to make the holes for the PMT seating and LEDs. Later, an acrylic coating is applied to the machined PU matrix to ensure the gel curing process remains uninterrupted.

\begin{figure}[!htb]
\begin{center} \includegraphics[scale=0.19]{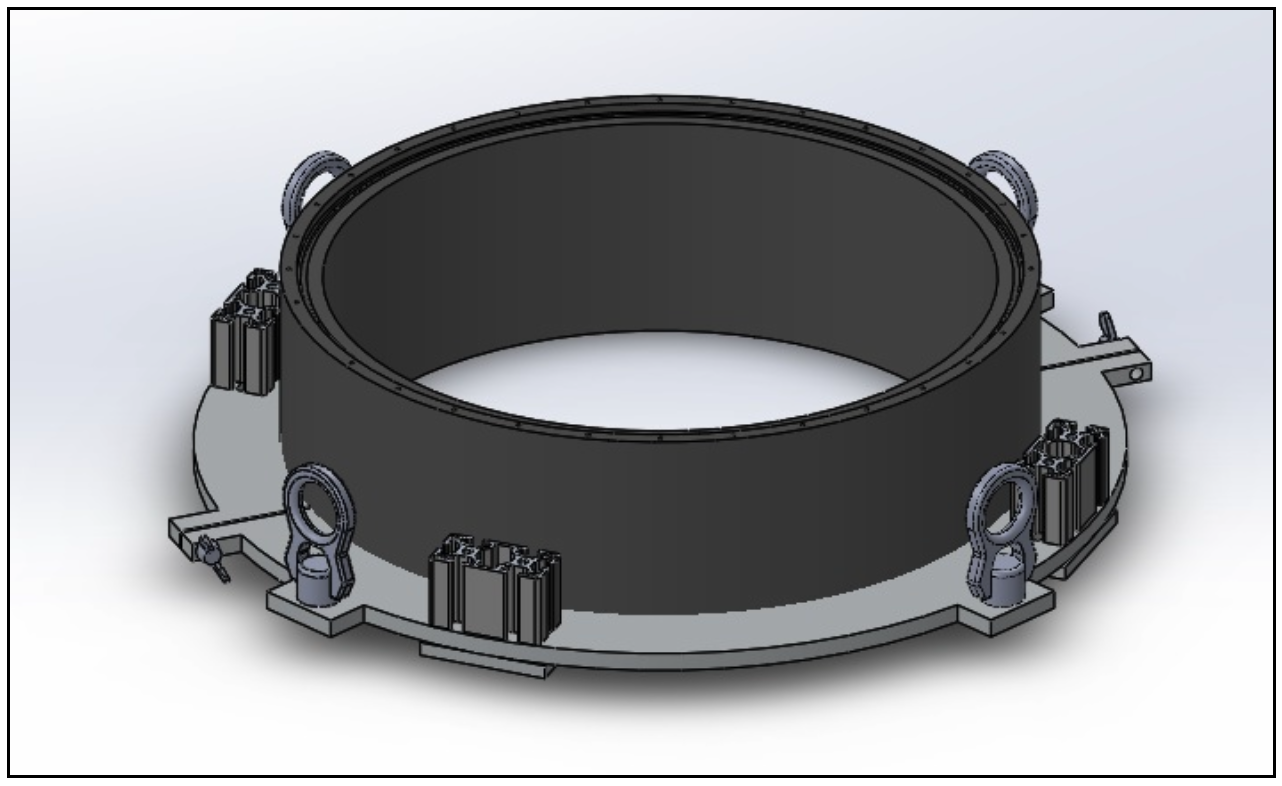} \end{center}
\vspace{-10pt}
\caption{A model of a PVC cylinder along with the assembly ring installed in the groove.}
\label{fig:PVC_Cylinder}
\end{figure}

\subsection{PVC Cylinder}
The body of the mPMT consists of a cylinder that houses the PMTs and electronics. We chose Polyvinyl Chloride (PVC) for these cylinders due to its low cost and minimal impact on the quality of ultra-pure water. The model of the PVC cylinder is shown in Figure~\,\ref{fig:PVC_Cylinder}. The cylinders are made from commercial 20-inch schedule-80 PVC pipes, cut to the required size for mPMT construction. The cylindrical shape was chosen primarily for its strength in withstanding water pressure. For ease of handling, a temporary assembly ring is placed around the cylinder in a dedicated groove, allowing for easy lifting. The PVC is machined at both ends to ensure a water-tight seal between the cylinder, the acrylic dome, and the backplate using O-rings. Stainless steel AISI304 Helicoil inserts are embedded in the PVC to hold components together and compress the O-rings for a watertight fit. Prototypes using these PVC cylinders were built in Canada and Poland to test the pressure tolerance of the acrylic dome, clamp ring, and cable penetrators.

\begin{figure}[!htb]
\begin{center} \includegraphics[scale=0.29]{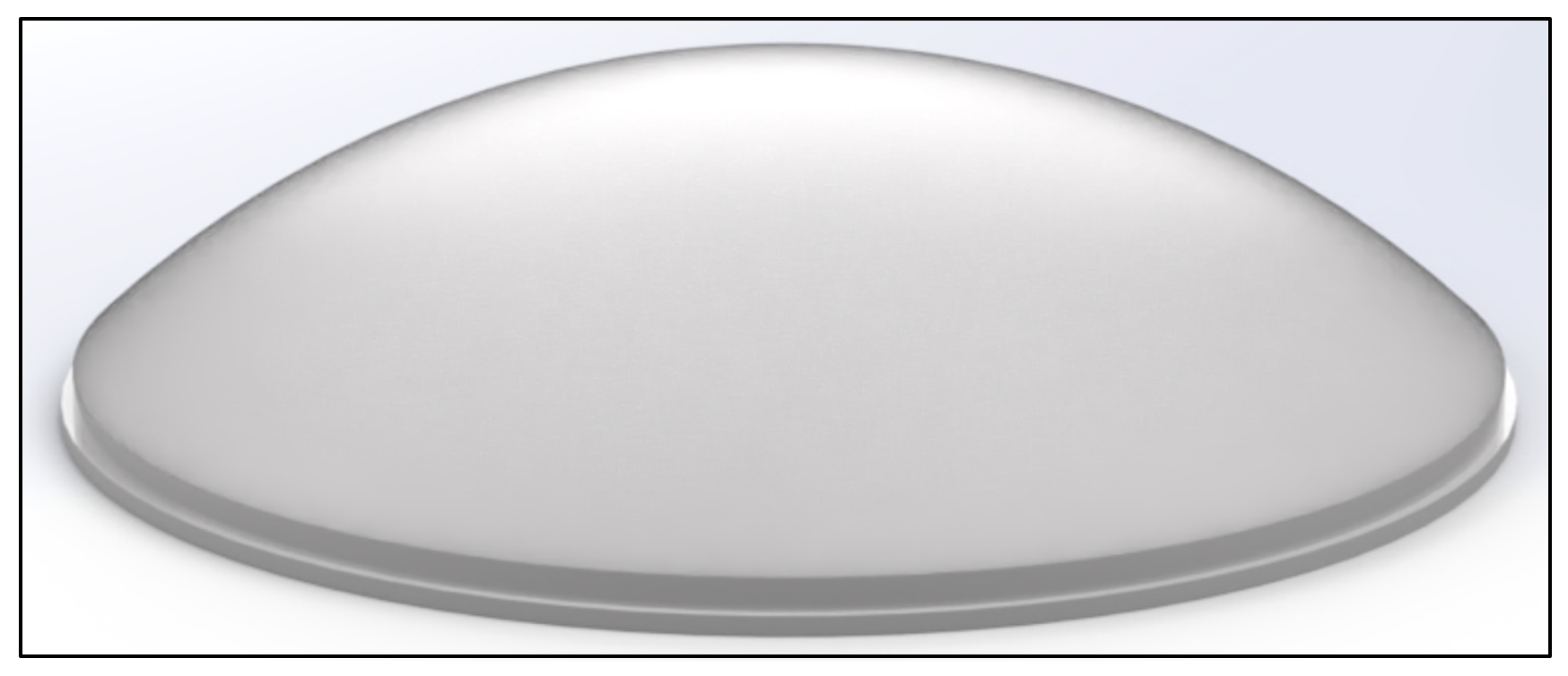} \end{center}
\vspace{-10pt}
\caption{A model of the acrylic dome.}
\label{fig:Acrylic_Dome}
\end{figure}

\subsection{Acrylic Dome}

Both the experience gained from the Super-K experiment and the R\&D done for the Hyper-K 20-inch PMTs \cite{20-inch PMT} helped in finding the right material for the acrylic dome. Transmittance and reflectance measurements were done on several samples delivered by the manufacturers to choose the material with the best optical properties. The transmittance of the 15~$\rm mm$ thick samples was measured in water and air, both with and without a 5~$\rm mm$ layer of optical gel. The Plexiglas GS UVT acrylic by Roehm GmbH\footnote{Roehm: \url{https://www.roehm.com/en/}} was chosen based on its properties. The production method for the dome is as follows: Roehm GmbH produces flat Plexiglas 3 $\rm m$~$\times$~2 $\rm m$ sheets; Liras SRL then thermoforms and machines the domes to the specified dimensions. The picture of the model of the dome is shown in Figure~\,\ref{fig:Acrylic_Dome}. After receiving the domes from Liras the shape of their inner surface is measured using a ROMER Absolute Arm 6-axis scanner. The domes have a thickness of 15~$\rm mm$ and we observed a maximum deformation of 0.5~$\rm mm$ along the radial direction when 1.5~$\rm MPa$ pressure was applied to the dome when installed into the mPMT. 

\begin{figure}
\begin{center} \includegraphics[scale=0.22]{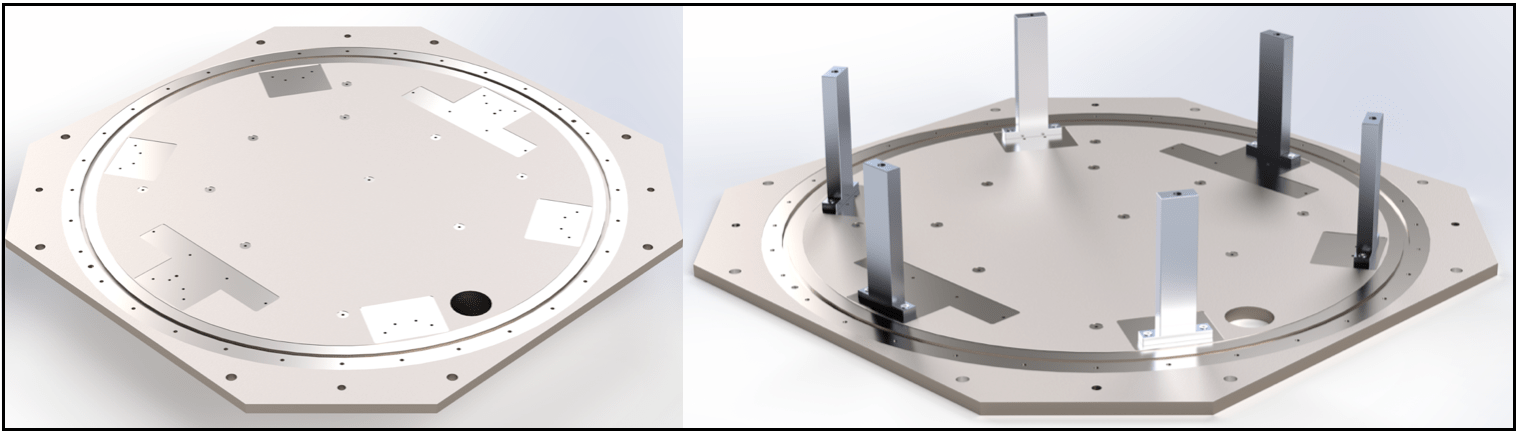} \end{center}
\caption{A model of the stainless steel backplate with/without stainless steel pillars.}
\label{fig:Stainless_Steel_Backplate}
\end{figure}

\subsection{Stainless Steel Backplate and Pillars}
The backplate is comprised of a machined three-eighth-inch (3/8-inch) plate of stainless steel AISI304. The top surface has skimmed areas to ensure proper seating between the PVC cylinder and aluminium pillars, and there are thru-holes and tapped holes to allow for the assembly and mounting of the mPMTs in the WCTE support structure. The design of the stainless steel backplate with the pillars is shown in Figure~\,\ref{fig:Stainless_Steel_Backplate}. The support pillars are designed to ensure the proper seating of the PMT support matrix on the top and attached to the tapped holes of the backplate. The pillars have a tapped hole on the top which is used to secure the PMT support matrix. The backplate also includes a separate hole (pressure cap hole) that could be used to apply a vacuum for better contact between the gel and acrylic, and a recessed groove designed to fit an O-ring matching the size of the cylinder. The backplate is integrated with a feed-through, designed to connect the onboard mPMT electronics to the exterior while ensuring water-tightness. 

\begin{figure}
\begin{center} 
\includegraphics[scale=0.21]{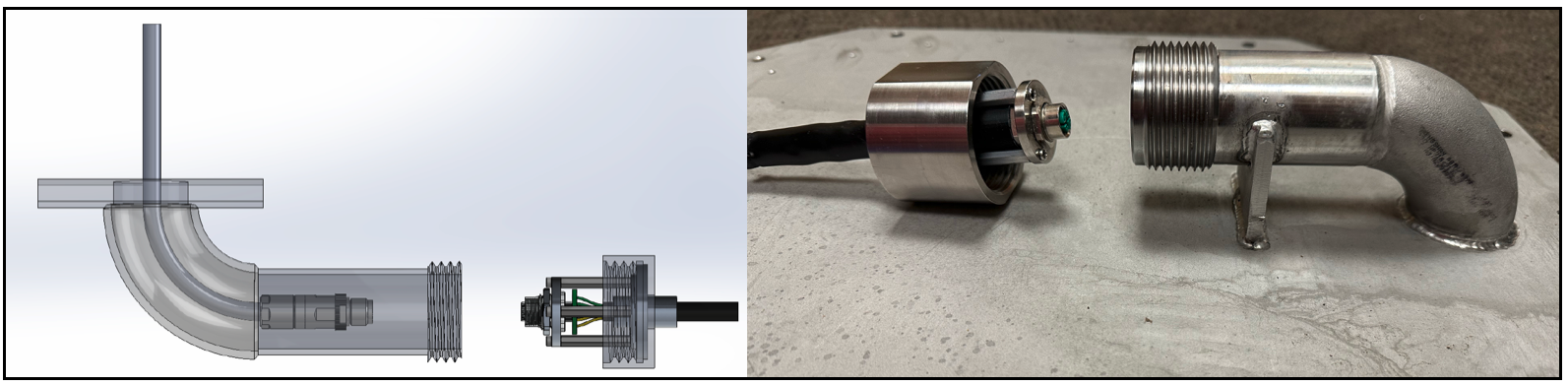}
\end{center}
\vspace{-10pt}
\caption{(a) Design of the stainless steel feedthrough (left) and (b) actual feedthrough prototype (right).}
\label{fig:Feed_through}
\end{figure}

The feed-through, shown in Figure~\,\ref{fig:Feed_through}, features a stainless steel elbow and flange welded to the backplate and a flange pre-attached to a water-tight cable with silicone, epoxy, and waterproof heat shrink. An M12 network connector is used for greater reliability compared to RJ45. The outer flange cable connects to a short network cable within the mPMT, and the outer flange cover is screwed onto the inner flange, compressing an O-ring to ensure a secure watertight seal. 


\section{mPMT Assembly Procedure}\label {s:mPMT Assembly Procedure}

\subsection{Ex-Situ PMTs Gelling Procedure}
R14373-PMTs, manufactured by Hamamatsu Photonics\footnote{Hamamatsu Photonics: \url{https://www.hamamatsu.com/jp/en/product/optical-sensors/pmt.html}} are used in the construction of mPMTs. Each PMT is separately covered by an optical gel before it is used in the mPMT ex-situ assembly. This procedure is called the ex-situ gelling procedure. In this procedure, the PMTs are prepared and stored separately from the other mPMT components. These PMTs have been characterised and tested for compliance with the Hyper-K requirements. Upon receiving the PMTs from Hamamatsu the High-Voltage/Front-End (HV/FE) boards are installed on each PMT and before gelling each PMT is tested to check the functionality, dark rate, etc.  The ELASTOSIL RT 604 A/B two-component silicon gel made by WACKER\footnote{WACKER: \url{https://www.wacker.com/h/en-us/silicone-rubber/room-temperature-curing-silicone-rubber-rtv-2/elastosil-rt-604-ab/p/000009552}} has been used for the gelling of the PMTs because of its fast curing, transparency, and viscosity (800~$\rm mPa.s$). Extensive R\&D has been performed to understand the mixing ratio of the two components, mixing time, and degassing time to make sure the gel is not unnecessarily sticky and give better contact with the acrylic dome cover.  Finally, the 96:4/A:B gel ratio is mixed using a pneumatic paint shaker for three minutes, and then degassed for five minutes.

 \begin{figure}[!htb]
\begin{center} \includegraphics[scale=0.38]{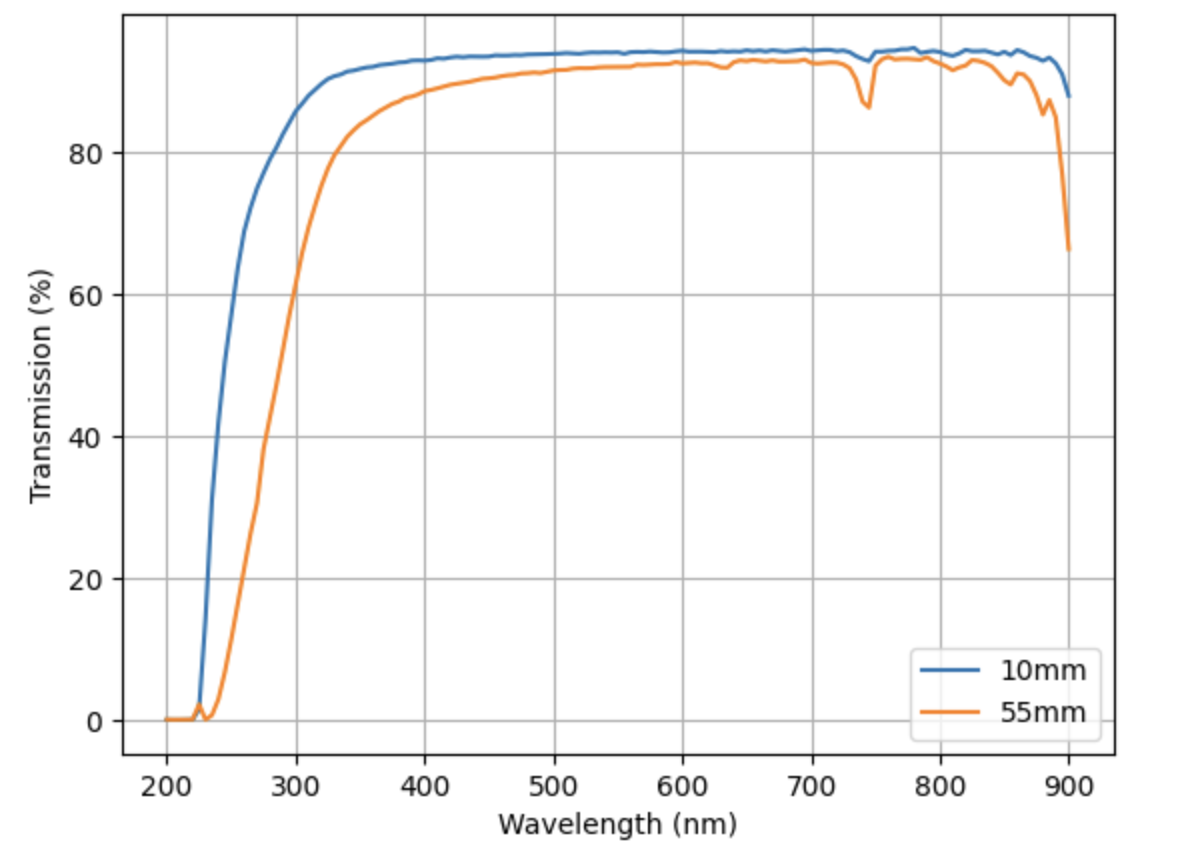} \end{center}
\vspace{-10pt}
\caption{Transmission measurements for varying thicknesses of ELASTOSIL RT 604 gel in 96:4 ratio. To improve readability, only the transmission probabilities for the minimum and maximum gel thicknesses are plotted.}
\label{fig:Gel_Transmission}
\end{figure}

Figure~\,\ref{fig:Gel_Transmission} shows the transmission measurements for several gel thicknesses after curing done using a Carry-60 spectrometer. As one might expect, thicker samples transmit less light across all wavelengths due to increased absorption within the material. The largest difference between thicknesses is present below 300~$\rm nm$, however, the PMT is not very sensitive in the below 300nm region. Hence this various thickness of the gel does not play a significant role in mPMT performance.

\begin{figure}
\begin{center} \includegraphics[scale=0.28]{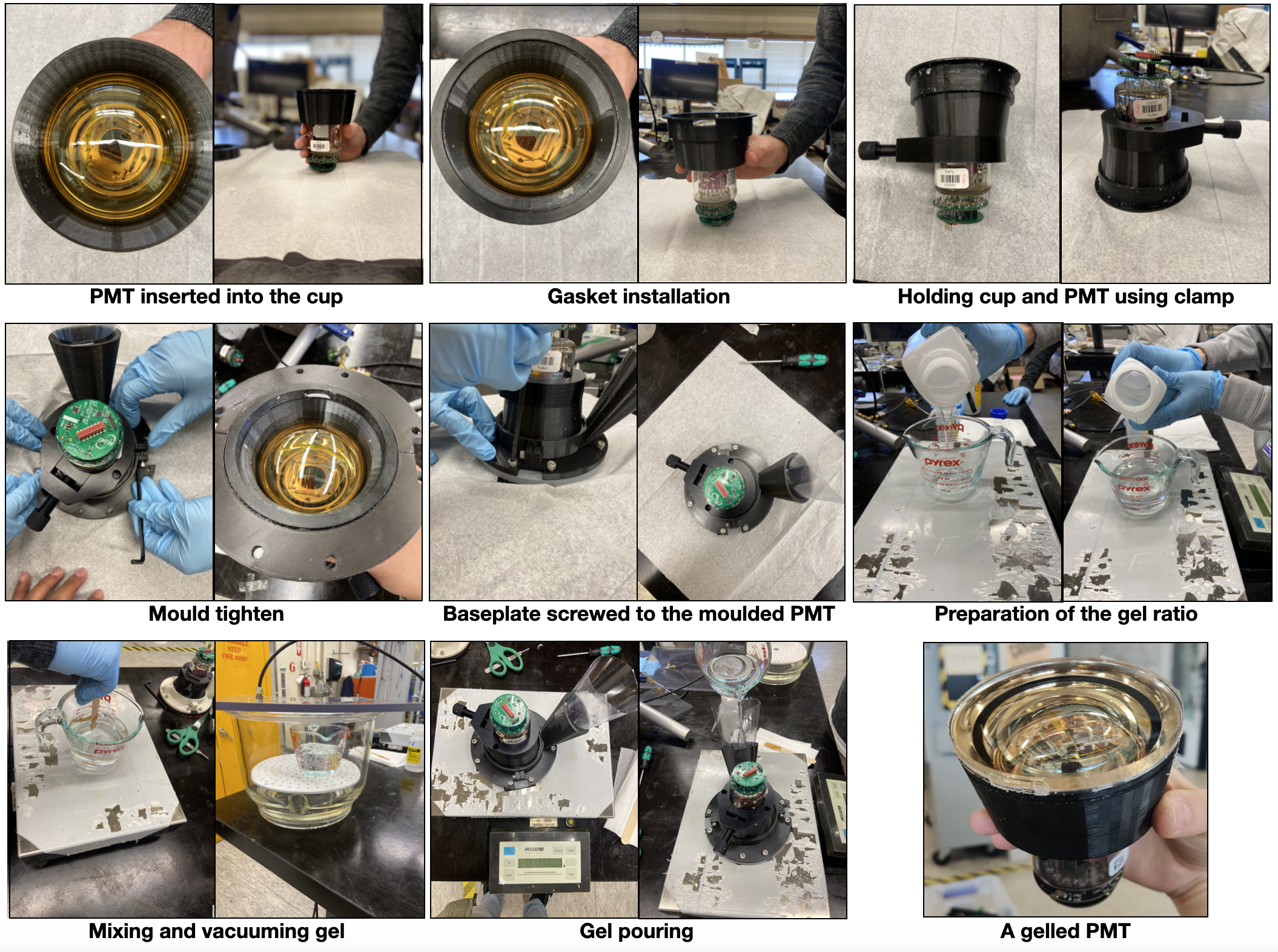} \end{center}
\vspace{-10pt}
\caption{A step-by-step procedure to prepare and gel an ex-situ PMT.}
\label{fig:Gelling_Procedure}
\end{figure}

Several iterations were done in designing the gelling tools to refine the procedure for the gelling and the cleaning of the components after gelling such that the total amount of time for preparation and cleaning of the components should be at most 10-15 minutes for each PMT. The mechanical components involved in the assembly of the PMT for gelling using the ex-situ gelling scheme consist of a machined baseplate, a 3D-printed two-piece mould, a cup, a gasket, a clamp, and a plastic funnel.  The machined mould is surrounded by an O-ring and capped with an acrylic plate with the same curvature radius as the acrylic dome, ensuring no air gap between the dome and the gel. Figure~\,\ref{fig:Gelling_Procedure} highlights the major steps involved in the gelling of a PMT and the final product i.e. a gelled PMT. The assembly procedure for the gelling of an ex-situ PMT is as follows: The PMT is inserted into a cup along with the gasket which helps to control the gel shape and height. The reflector ring is installed on the lip of the 3D-printed cup. Then cup+PMT+gasket is held in the mould and screwed down to the baseplate. The gel is poured through the funnel at the bottom of the structure and placed in an incubator at $35^\circ\mathrm{C}$ for 24 hours to allow curing to take place. A total of approximately 2.2~$\rm kg$ of gel is required to prepare 20 PMTs at a time in this procedure.

\begin{figure}
\begin{center} 
\includegraphics[scale=0.17]{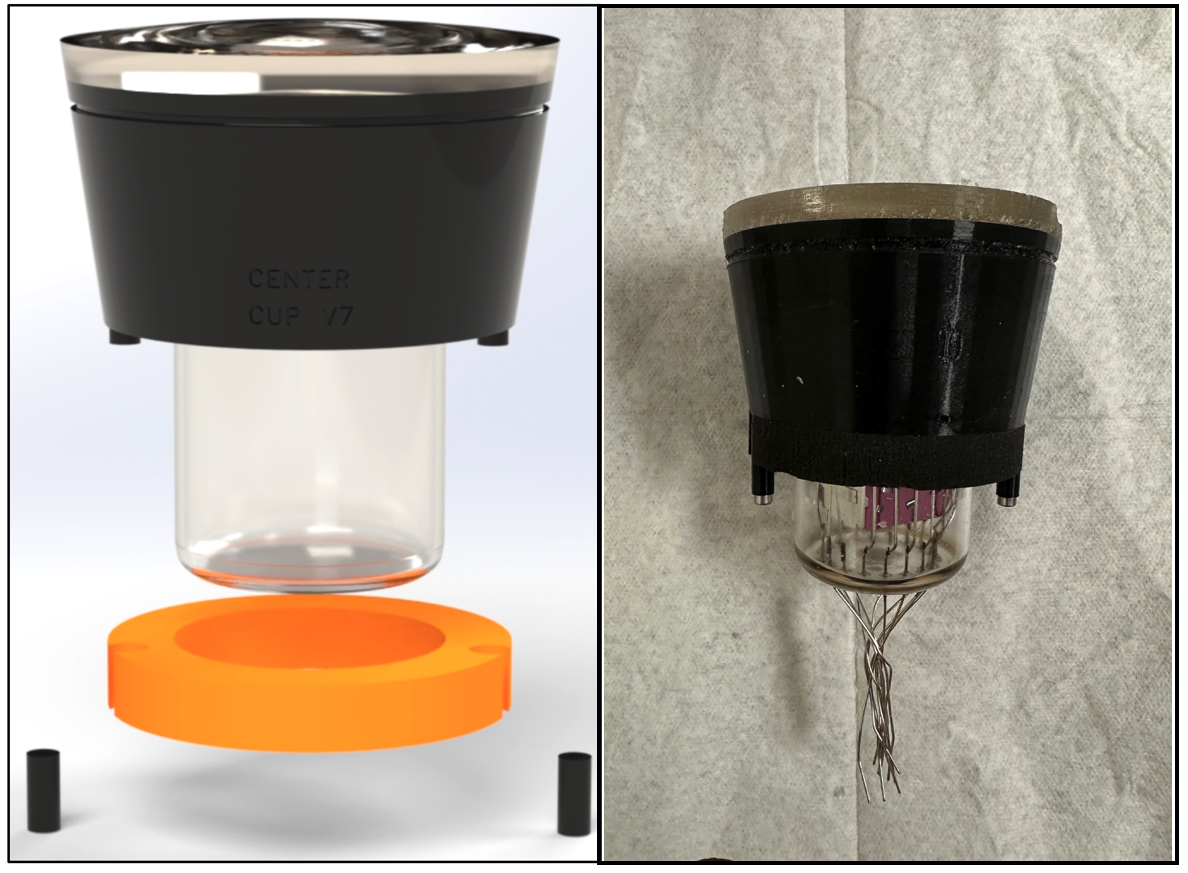}
\end{center}
\vspace{-10pt}
\caption{(a) Design for the installation of the Poron ring and seating pins on an ex-situ gelled PMT (left), and (b) actual PMT equipped with the Poron ring and seating pins (right). The poron is black in reality. Hence, we can't see the difference between poron and cup in the picture}
\label{fig:Poron_Installation}
\end{figure}

\subsection{Ex-Situ Assembly Procedure}
\begin{figure}
\begin{center} \includegraphics[scale=0.2]{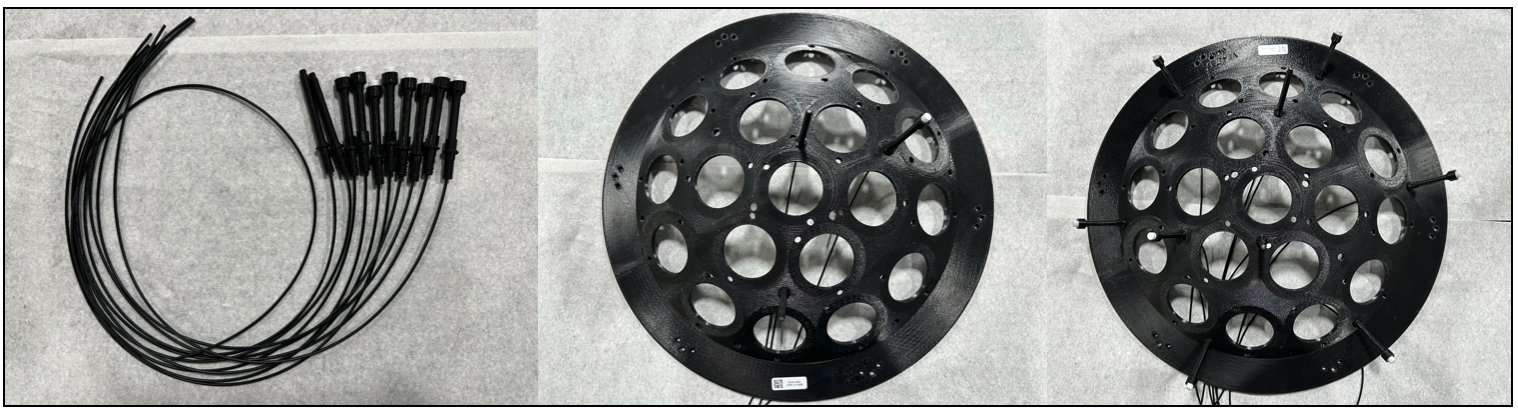} \end{center}
\vspace{-10pt}
\caption{The picture shows the preparation of the light pipes for the ex-situ matrix. The left panel shows the optical fibres attached to 3D-printed LED holders. Two types of holders are used: those with white caps correspond to diffused outputs, while the bare black holders correspond to 15-degree collimated outputs. The centre and right panels show the process of mounting these holders onto the matrix. In this configuration, seven holders are used for diffused light and two for collimated light.}
\label{fig:Ex-Situ_Matrix_Preparation}
\end{figure}

The assembly of an ex-situ mPMT photosensor is primarily divided into four stages: component preparation, backplate preparation, dome sub-assembly, and detector closing. Component preparation involves installing Poron\textsuperscript{\textregistered} (poron is trademarked name for a family of microcellular polyurethane foam) rings and seating pins onto the gelled PMTs, as shown in Figure~\,\ref{fig:Poron_Installation}, as well as fitting the light pipes into the designated holes in the matrix. As previously mentioned, each mPMT includes six light pipes with a diffused output for the photogrammetry target, along with three LEDs for detector calibration, one with a diffused output and two with a 15 or 30-degree collimated output, as shown in Figure~\,\ref{fig:Ex-Situ_Matrix_Preparation}. The following subsections provide more detailed descriptions of the dome sub-assembly, backplate preparation, and detector closing.

\subsubsection{Dome Sub-Assembly} \label{s:dome_assembly}
\begin{figure}
\begin{center} \includegraphics[scale=0.2]{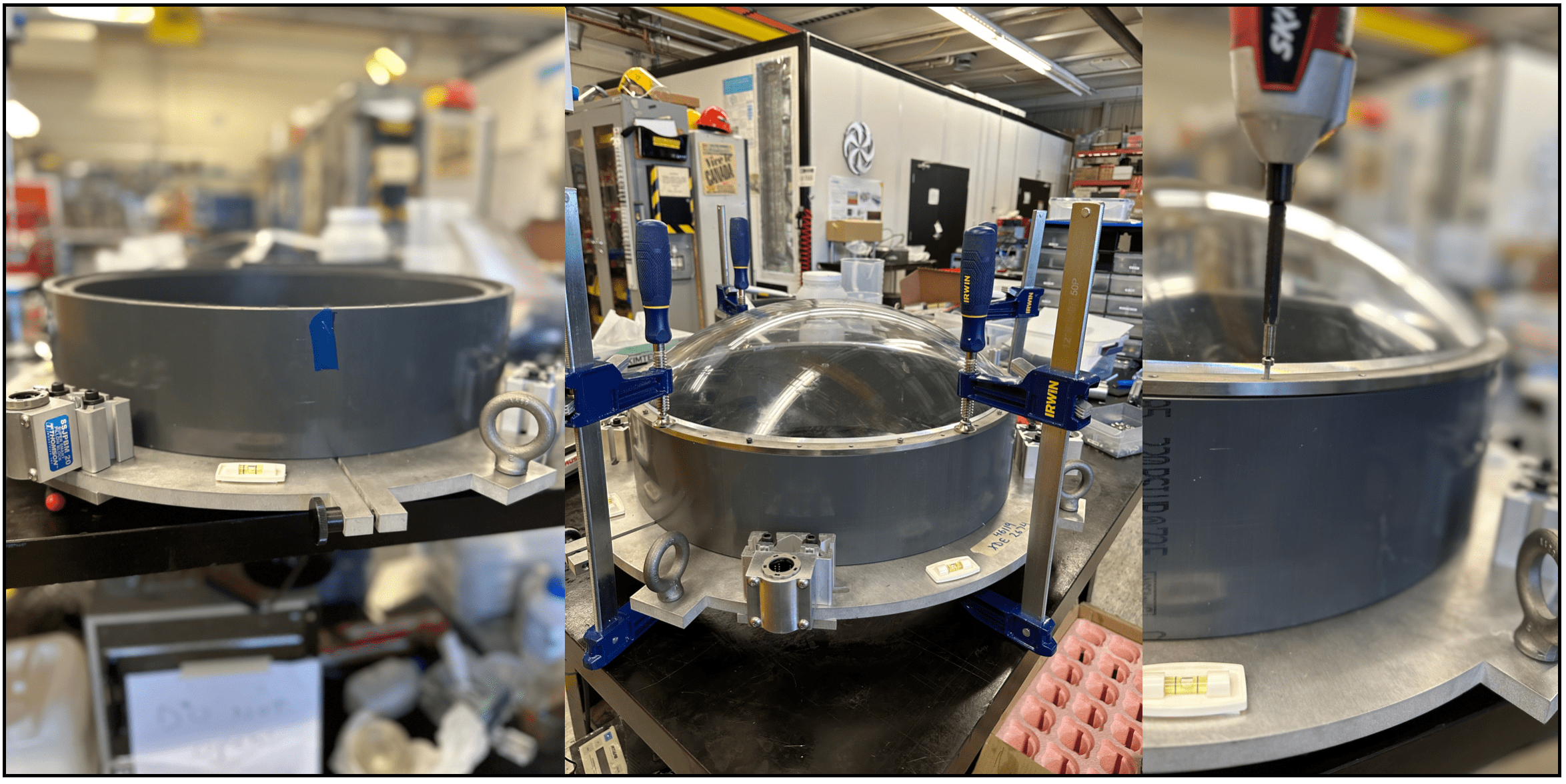} \end{center}
\vspace{-10pt}
\caption{The picture shows the dome sub-assembly. Installation of the assembly ring and the O-ring on the cylinder, covering the cylinder with the dome and attaching them with the clamp ring.}
\label{fig:Dome_Subassembly}
\end{figure}

Once all the PMTs are ready and tools are available, one can start the dome sub-assembly. Before starting the dome sub-assembly, it is important to install the assembly ring around the PVC cylinder, which helps handle the cylinder alone and lowers the dome sub-structure to achieve gradual contact with the gelled PMTs without damaging them. The dome sub-assembly includes fixing the dome to the PVC cylinder with the clamp ring as shown in Figure~\,\ref{fig:Dome_Subassembly}. The O-ring in the dedicated cylinder groove allows for a water-tight seal between the PVC and the acrylic dome.  The PVC cylinder has two dowel pin holes to help in the alignment of the mounting bolts on the cylinder as well as on the backplate. The final step in the preparation of the dome substructure for mPMT assembly is cleaning the inner surface of the dome and cylinder rigorously with isopropyl alcohol to get rid of any dust present.

\vspace{-10pt}

\subsubsection{Backplate Preparation}
Preparing the backplate involves installing the electronics, connecting the flat ribbon cables to the frontend board at the base of the PMTs, and securing the pillars along with the matrix on top of them. The backplate is resting on the assembly jig, which is specifically designed to facilitate the precise lowering of the dome substructure.

\begin{figure}
\begin{center} \includegraphics[scale=0.24]{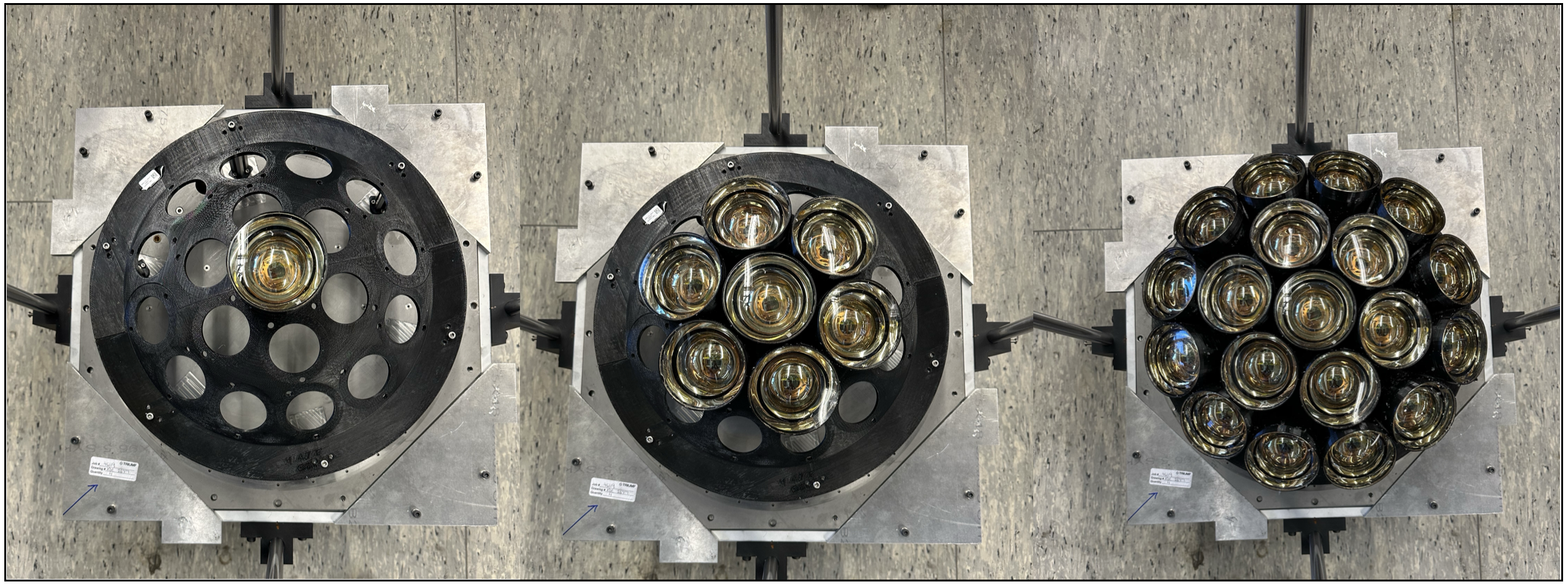} \end{center}
\vspace{-10pt}
\caption{Installation of PMTs into the inner support matrix before closing the detector.}
\label{fig:PMTs_Installation}
\end{figure}

It is constructed from aluminium extrusions and features an octagonal cutout on top that perfectly accommodates the backplate. Four guide rods are positioned at the centre of each side of the jig, ensuring the dome substructure, which has an assembly ring with bearings fitted into a cylinder groove, can smoothly and accurately roll along the rods during the lowering process. With the backplate positioned on the jig, the mainboard is installed onto it, and flat ribbon cables are connected to each channel of the mainboard. The matrix is placed on top of these pillars, and the PMTs are installed one by one into the designated holes and attaching the flat ribbon cable to the bottom of the PMT according to the desired mapping between the mainboard channels and the PMT position as shown in Figure~\,\ref{fig:PMTs_Installation}.

\subsubsection{Detector Closing}

\begin{figure}
\begin{center} \includegraphics[scale=0.25]{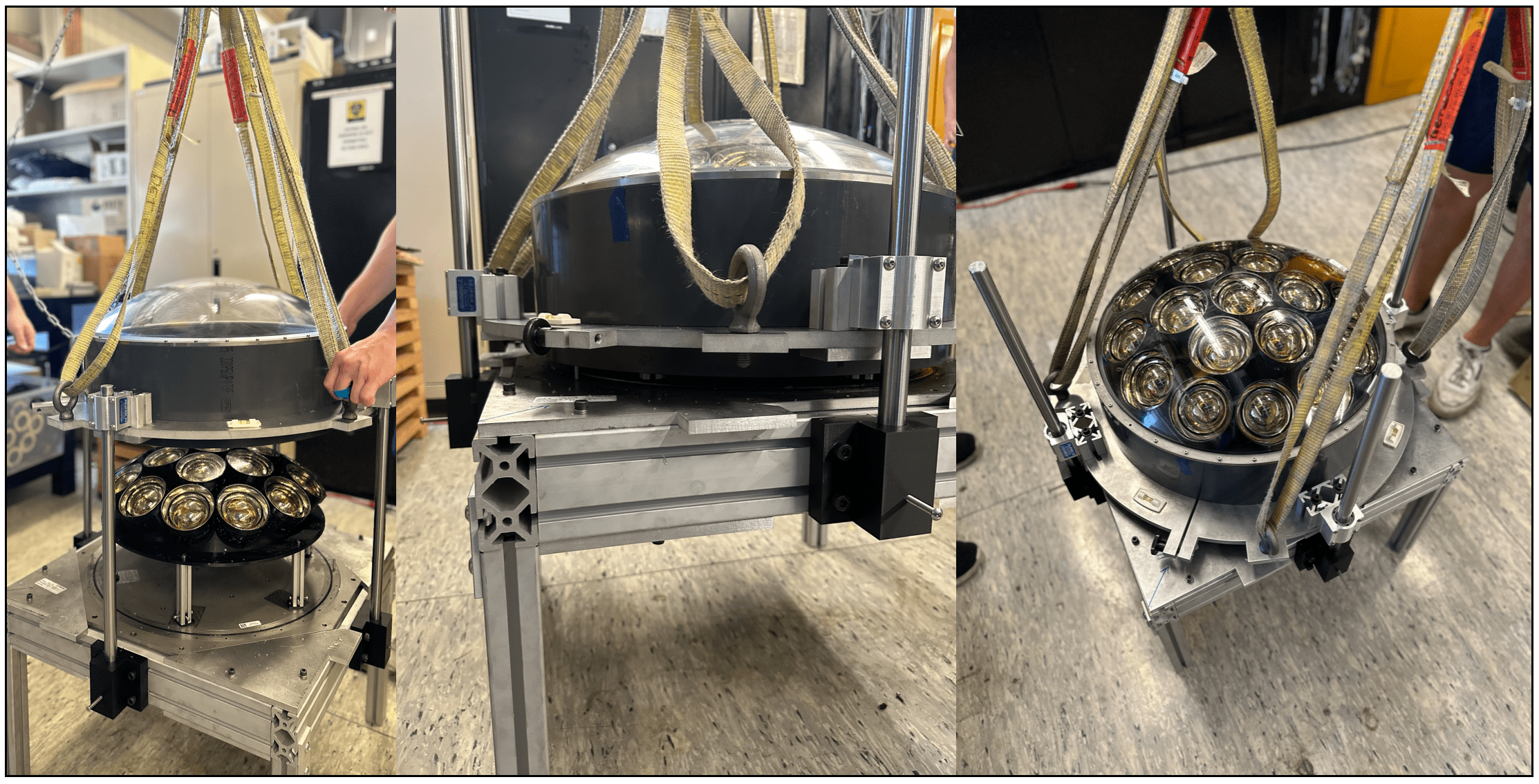} \end{center}
\caption{Lowering the dome sub-assembly structure onto the backplate having PMTs installed on the inner support matrix.}
\label{fig:Lower_Dome}
\end{figure}

\begin{figure}
\begin{center} \includegraphics[scale=0.4]{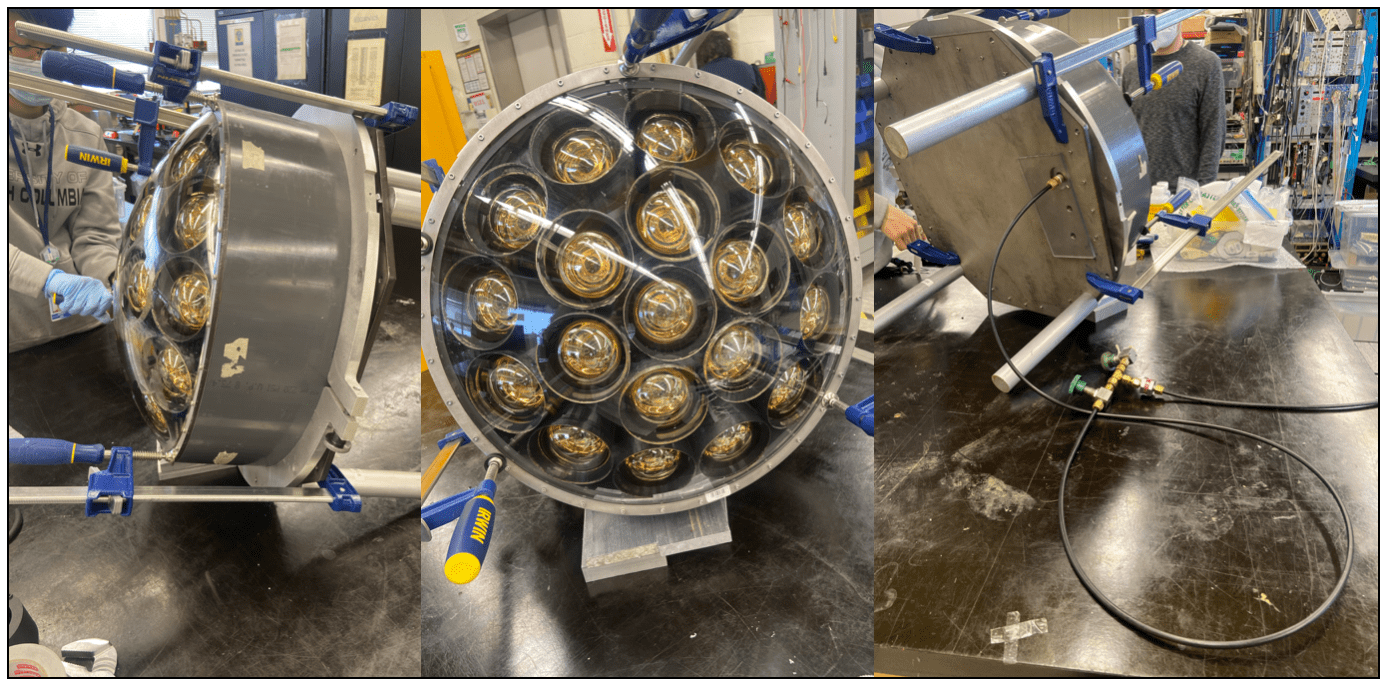} \end{center}
\caption{Once the backplate is securely screwed to the cylinder, begin applying the vacuum from outside the detector through the pressure cap hole.}
\label{fig:Applying_Vacuum}
\end{figure}

Before the final assembly of the mPMT, the PMTs undergo a cleaning process. The surfaces of the gel on the PMTs are carefully wiped with isopropyl alcohol using wipes, followed by a visual inspection to ensure the removal of any residual dust. If necessary, the cleaning is repeated until all surfaces are contaminant-free. Once dry, a silicon spray is applied to each PMT to ensure optimal optical contact with the dome without shearing the gel. The dome sub-assembly is then carefully lowered onto the backplate. During this process, care is taken to avoid contact between the cylinder and the PMT surfaces, as illustrated in Figure~\ref{fig:Lower_Dome}. The dowel pins on the backplate are aligned with the corresponding holes in the cylinder. At this stage, the cylinder is clamped at four points along its perimeter to facilitate screwing from the bottom of the backplate. The cylinder is secured using screws from the backside of the backplate. With the assembly of the mPMT detector complete, a vacuum is applied to the detector for at least two hours, maintaining a pressure of approximately 40\% of atmospheric pressure. This is achieved by connecting a suction pipe to the pressure cap hole on the backside of the plate, as depicted in Figure~\,\ref{fig:Applying_Vacuum}.. The vacuum serves to remove any trapped air between the gel and the dome, ensuring optimal contact. Once full contact is established, the system is returned to atmospheric pressure by turning off the vacuum.

\subsection{In-Situ Assembly Procedure}\label {s:in-situ mPMT Assembly Procedure}

The in-situ mPMT assembly procedure does not require gelling the PMTs beforehand; this makes it different from the ex-situ assembly procedure. An in-situ mPMT assembly is also divided into four parts: components preparation, backplate preparation, dome sub-assembly, and detector closing. 

\begin{figure}
\begin{center} \includegraphics[scale=0.3]{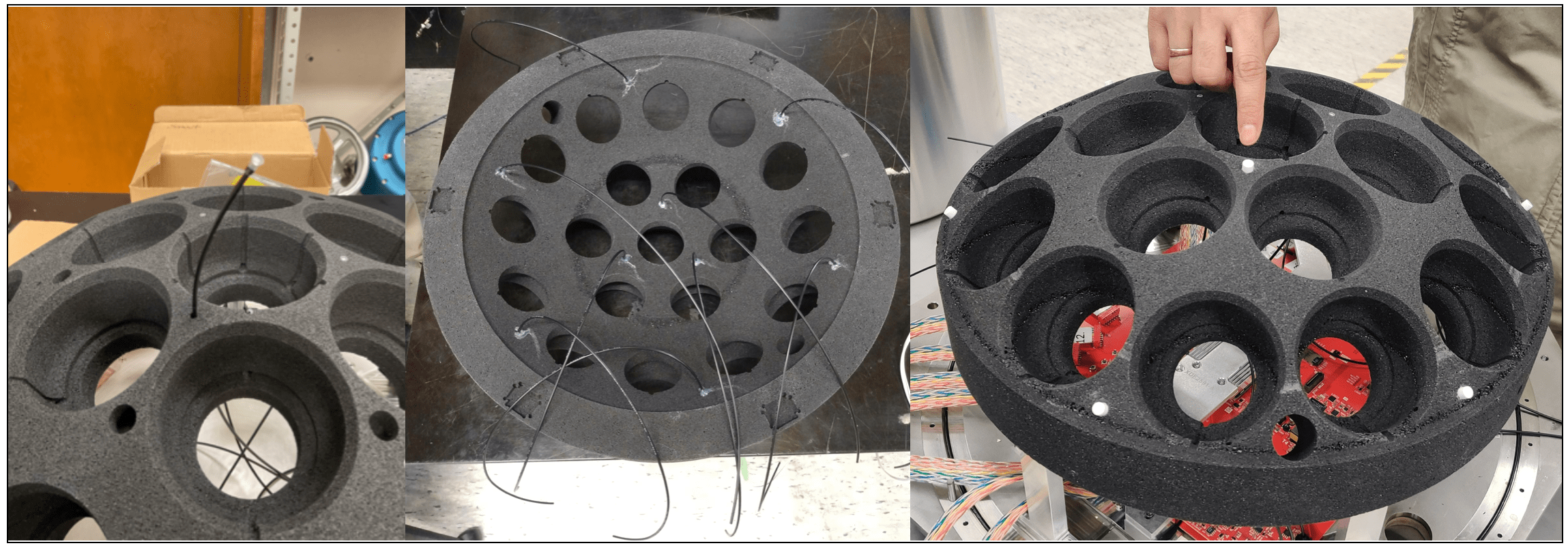} \end{center}
\caption{Installation of the light pipes as well as diffuser caps on the dedicated holes on the matrix.}
\label{fig:In_Situ_Matrix_Preparation}
\end{figure}

The components preparation involves the preparation of the matrix, which includes glueing the light pipes into the dedicated holes in the matrix and the diffuser cap, as shown in Figure~\,\ref{fig:In_Situ_Matrix_Preparation}.  The dome substructure assembly procedure for the in-situ mPMT construction is similar to the ex-situ procedure as described in Section~\ref{s:dome_assembly}. The following sections provide a more detailed description of the backplate preparation, gel pouring and the detector closing process.

\subsubsection{Backplate Preparation}
The preparation of the backplate for the in-situ mPMT differs slightly from that of the ex-situ mPMT. Notably, the PMT support matrix used in the in-situ mPMT is different from the one used in the ex-situ, as detailed in Section~\ref{s:matrix}. The backplate preparation involves installing the electronics, attaching the flat ribbon cables, securing the pillars, and glueing the matrix onto the pillars instead of screwing.

\begin{figure}
\begin{center} \includegraphics[scale=0.3]{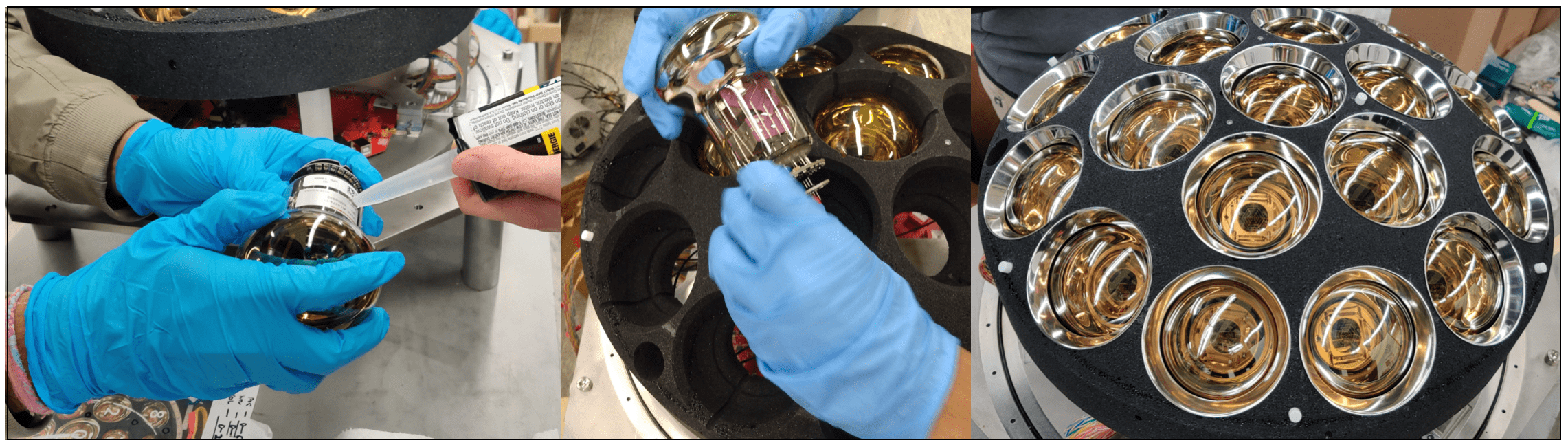} \end{center}
\caption{Installation of the PMTs on the dedicated holes on the matrix.}
\label{fig:In_Situ_PMT_Installation}
\end{figure}

Beginning with the central PMT hole in the matrix, we apply a ring of silicone sealant on the tube part of the PMT to secure it in the matrix. Before placing the PMT in the matrix hole we attach the flat ribbon cable to the bottom of the PMT according to the desired mapping between the mainboard channels and the PMT position and glue the reflectors on the lip of each PMT hole on the matrix as shown in Figure~\,\ref{fig:In_Situ_PMT_Installation}. 

\begin{figure}
\begin{center} \includegraphics[scale=0.3]{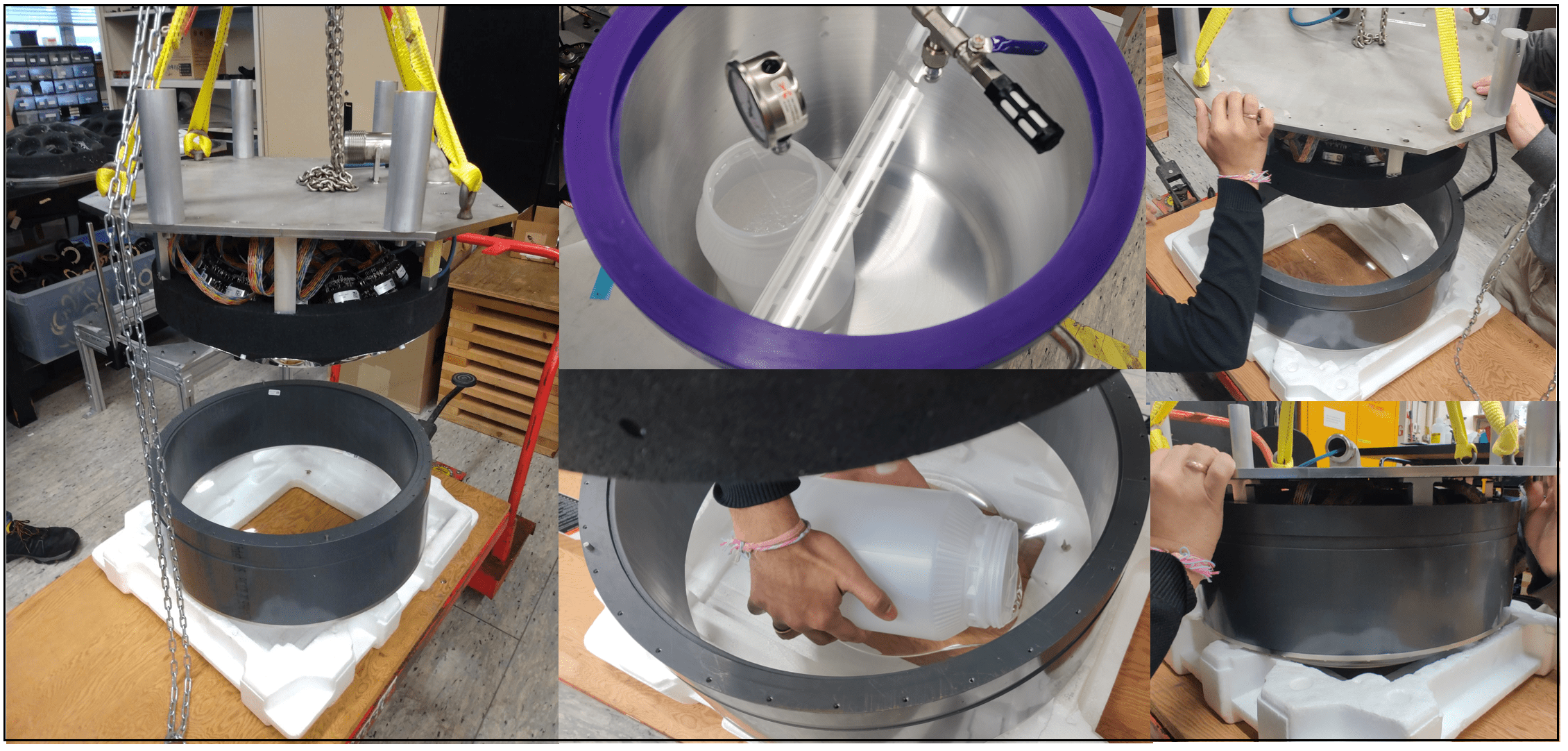} \end{center}
\caption{The image shows the de-gassing and pouring of the gel, lowering of the prepared backplate on the dome, and closing of the detector.}
\label{fig:In_Situ_Detector_Closing}
\end{figure}

\subsubsection{Gel Pouring and Detector Closing}
Verifying the fit of the components through a dry assembly is an essential step. In this process, the backplate is flipped upside down and attached to a crane. The dome substructure is also placed upside down, and the fully prepared backplate is then carefully lowered into position within the structure. Once the fit is confirmed, the backplate is removed using an overhead crane to facilitate the pouring of the gel. A total of 2.4~$\rm kg$ of gel is mixed, degassed, and poured into the dome, which is positioned on the assembly jig. The upside-down PMT on the backplate substructure is then lowered into the gel, and the detector is sealed by securing it from the backside of the backplate, as illustrated in Figure~\ref{fig:In_Situ_Detector_Closing}. Inserting the prepared matrix into the poured gel forces the gel to fill all the spaces between the PMTs and the dome. To prevent air pockets, air channels have been strategically added to the matrix, allowing air to escape. A level meter is employed on the backplate to monitor and correct any tilting, ensuring uniform gel distribution over the PMTs. This process completes the closure of the in-situ mPMT detector.  The mPMT is then left untouched for 1-2 days to allow the gel to fully cure.

\begin{table}[htb]
 \begin{adjustbox}{width=\textwidth}
  \begin{tabular}{ | c | c | c | c | c | c | } \hline
  Tasks & Ex-Situ Time (hr) & In-Situ Time (hr) & FTE Hours (ex-situ) & FTE Hours (in-situ) & Remarks \\ 
  \hline \hline
   Backplate Preparation & \ 1 & \ 1 \ & \ 2 \ & \ 2 \ & 2 FTEs \\ \hline
   PMT Gelling/PMT Gluing & \ 2 & \ 1 \ & \ 6 \ & \ 3 \ & 3 FTEs\\ \hline
   Detector Closing & \ 1.5 & \ 1.5 \ & \ 3 \ & \ 3 \ & 2 FTEs \\ \hline
   Moulds Cleaning & \ 2 & \ NA \ & \ 6 \ & \ NA \ &3 FTEs \\ \hline
   Total Time & \ 6.5 & \ 3.5 \ & \ 17 \ & \ 8 \ & 2-3 FTEs \\ \hline
   \end{tabular}
     \end{adjustbox}
   \caption{The actual time required to perform each task for both ex-situ and in-situ style per mPMT construction.}
   \label{tab:Time_Estimate}
\end{table}

\subsection{Time Estimates for In-situ Vs Ex-situ mPMT Assembly}

Based on our experience with WCTE mPMT production, we have estimated the total time and number of full-time equivalents (FTEs) hours required for each task involved in building both ex-situ and in-situ style mPMTs, which are shown in Table~\ref{tab:Time_Estimate}. The difference between the Ex/In-situ time and FTE hours is that total time refers to the total duration a task took, while FTE hours reflect how many personnel worked on that task. As can be seen, the time and FTE hours are lower for the in-situ assembly scheme. In summary, both the ex-situ and in-situ assembly strategies have been successfully implemented for mPMT construction. In both approaches, the optical gel uniformly fills the space between the PMTs and the acrylic dome, with no visible air gaps or bubbles observed. However, the in-situ method offers a clear advantage in terms of reduced assembly time and lower manpower requirements, making it the preferred approach for large-scale mPMT production.
\begin{figure}
\begin{center} \includegraphics[scale=0.35]{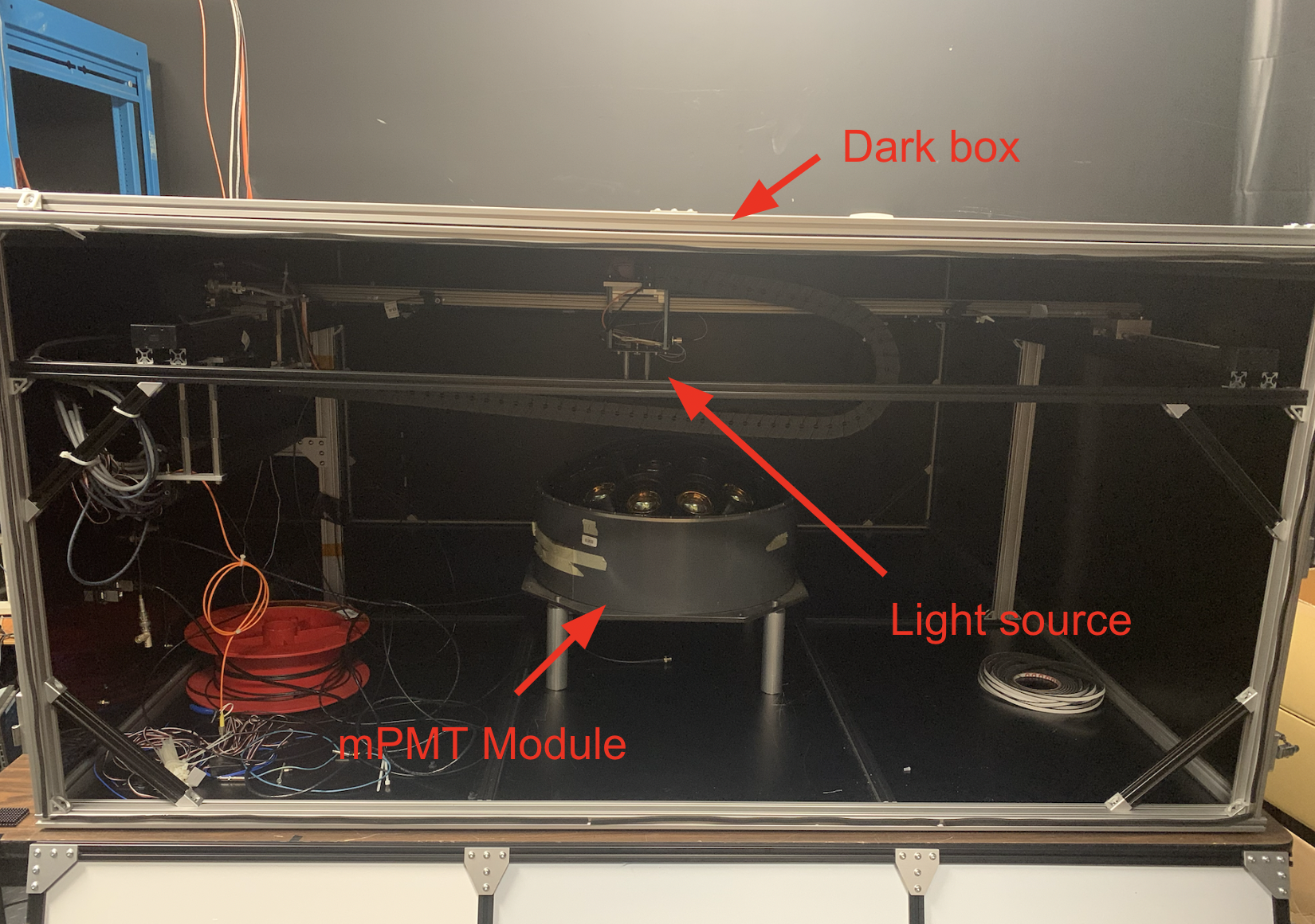} \end{center}
\caption{The mPMT test stand setup at TRIUMF.}
\label{fig:MTS}
\end{figure}

\section{mPMT Test Stand}\label{s:MTS}

The mPMT test stand (MTS) is a setup used for optical testing of the completed mPMT modules at TRIUMF. It comprises a dark box housing a light source attached to a gantry system. The light source used was a Tamadenshi LDB-200 series laser. It had a wavelength of 405 nm, and a pulse width between 40 and 80 picoseconds. The intensity was chosen such that a pulse was seen in a PMT receiving light approximately 10-30\% of the time since this ensures that most pulses are at the single photo-electron (1PE) level. This system enables movement in a two-dimensional coordinate system, allowing light to be pulsed at any point of interest within the dark box. A picture of the MTS can be seen in Figure~\,\ref{fig:MTS}. The test stand serves multiple purposes. During the mass production phase of the mPMT modules, it verifies the proper functioning of each PMT within a module by pulsing light at each PMT and measuring parameters such as dark rate, current-voltage curves, pulse height distribution for each fast-output LED, and relative efficiency. Another use is to gain a better understanding of the mPMT’s light-collecting capabilities. Specifically, studying the light-collecting properties of the aluminium reflectors that fit on the inner surface of the PMT cups is of interest. 

\begin{figure}
\begin{center} \includegraphics[scale=0.2]{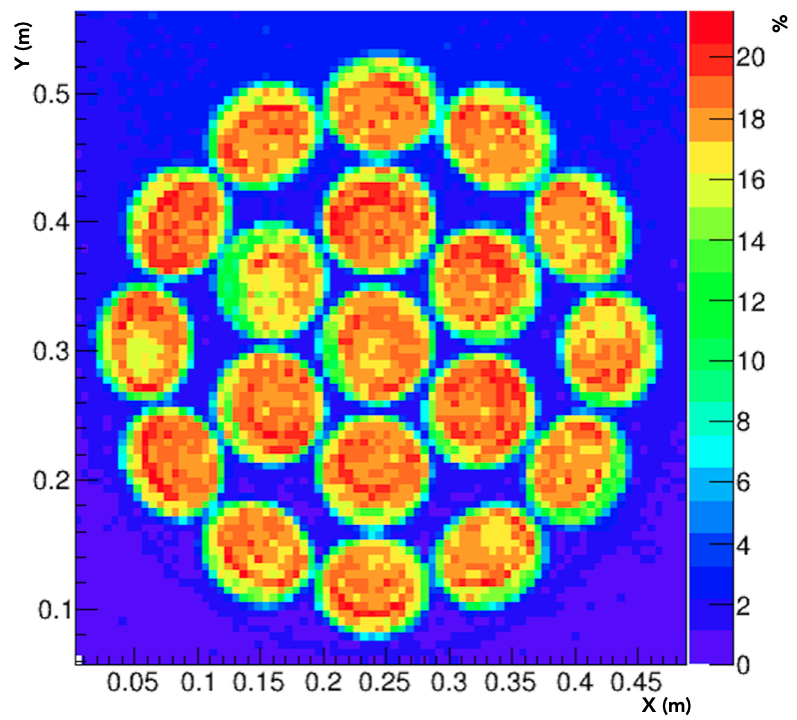} \end{center}
\caption{The summed efficiency plot for the mPMT module having all functional PMTs.}
\label{fig:mPMT_Summed_Efficiency}
\end{figure}

\subsection{mPMT Efficiency Measurement} 

A large number of mPMT modules have been assembled and tested at the MTS facility. The results from some particular mPMT modules containing all functioning PMTs with/without reflectors are discussed in the following sections. The single PMT relative detection efficiency plot is a histogram showing the number of laser pulses detected by the PMT divided by the total number of pulses collected for each given XY position of the light source. The summed efficiency plot is the result of plotting all individual detection efficiency plots from each separate PMT. An example of the summed efficiency plot for the module with reflectors installed on all PMTs is shown in Figure~\,\ref{fig:mPMT_Summed_Efficiency}.  The summed efficiency plot is simply the result of adding together all the individual PMT maps. So for each laser position (pixel), we (overlay) sum up the relative efficiency from all 19 PMTs. This gives us a single plot that shows how the full mPMT module responds to light at different positions, instead of showing 19 separate maps.


\begin{figure}
    \centering
    \begin{minipage}{0.45\textwidth}
        \centering
        \includegraphics[scale=0.183]{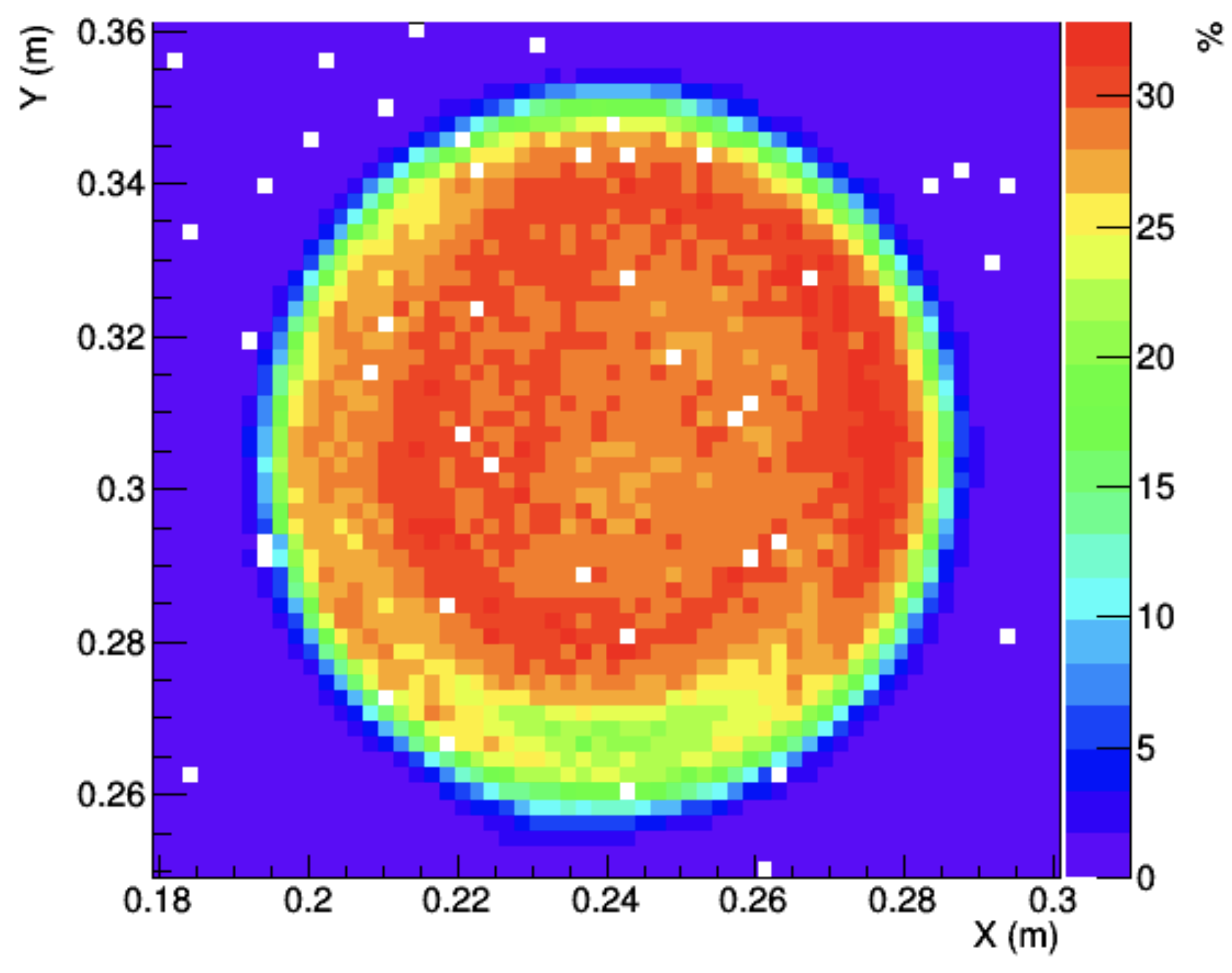}
        \caption*{(a) PMT with a reflector}
    \end{minipage}
    \hfill
    \begin{minipage}{0.45\textwidth}
        \centering
        \includegraphics[scale=0.18]{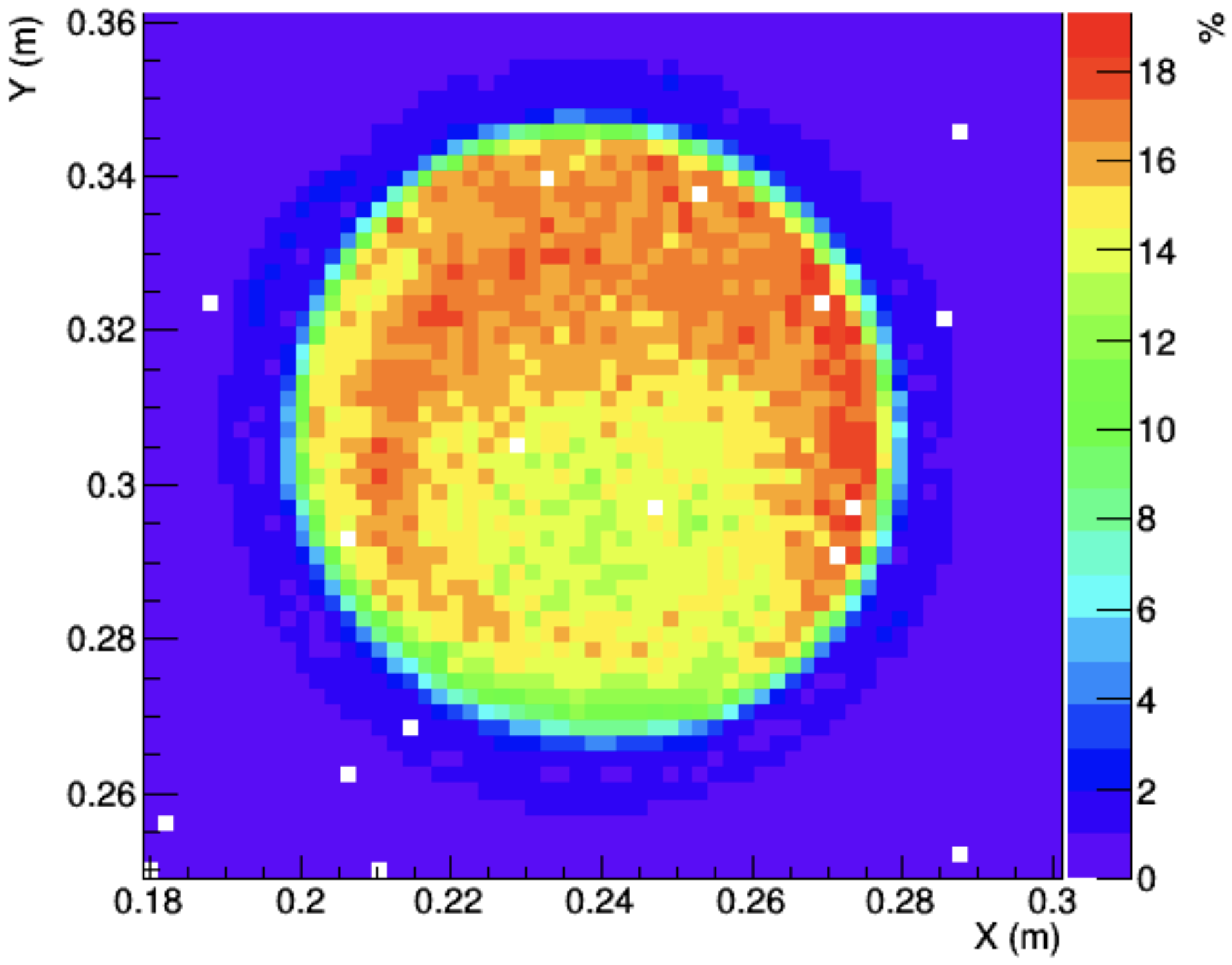}
        \caption*{(b) PMT without the reflector}
    \end{minipage}
    \caption{The plots show the efficiency measurement for the PMT at the center of the module: (a) with a reflector and (b) without the reflector.}
    \label{fig:Centre_PMT_Reflector_Efficiency_Comparison}
\end{figure}

\subsection{PMTs efficiency measurement with and without reflector}\label {s:Efficiency_Measurement_w/wo_Reflector}
It is important to understand the extent to which the effective photosensitive area of a PMT can be increased by adding a reflector around the PMT. To investigate this, a scan of several PMTs was conducted, and during the analysis process, the number of bins in each PMT efficiency histogram above the background threshold was counted. To find the background threshold, the average efficiency of bins that we don't expect to be in the PMT's photosensitive region was taken. The count of bins for each PMT was then converted into an \enquote{effective area}, representing the area from which each PMT is capable of detecting light.
A scan of the central PMT with a reflector is shown in Figure~\,\ref{fig:Centre_PMT_Reflector_Efficiency_Comparison}~(a). When counting the number of bins above the background threshold, the effective area is 66~$\rm cm^2$. The efficiency measurement for the central PMT from a separate module, which did not have a reflector, is illustrated in Figure~\,\ref{fig:Centre_PMT_Reflector_Efficiency_Comparison}~(b), with an effective area of 50~$\rm cm^2$. Therefore, the use of a reflector in this case has increased the photosensitive region by approximately 32\%. The observed difference in efficiency at the centre of the PMT, shown in~Figure~\,\ref{fig:Centre_PMT_Reflector_Efficiency_Comparison} (30\% (a) vs. 18\% (b)), is due to experimental difficulties and is not a result of the reflector used. This random fluctuation is mainly attributed to the differing laser intensity, as there was some intrinsic variation of the laser intensity between runs, which would impact the absolute efficiency. These are relative efficiency plots, so the efficiencies between runs should not be compared, but it is acceptable to compare the effective areas since the intensity should not affect this. The critical comparison should be based on the number of bins above the threshold, not the mean efficiency at the centre.


\begin{figure}
    \centering
    \begin{minipage}{0.45\textwidth}
        \centering
        \includegraphics[scale=0.19]{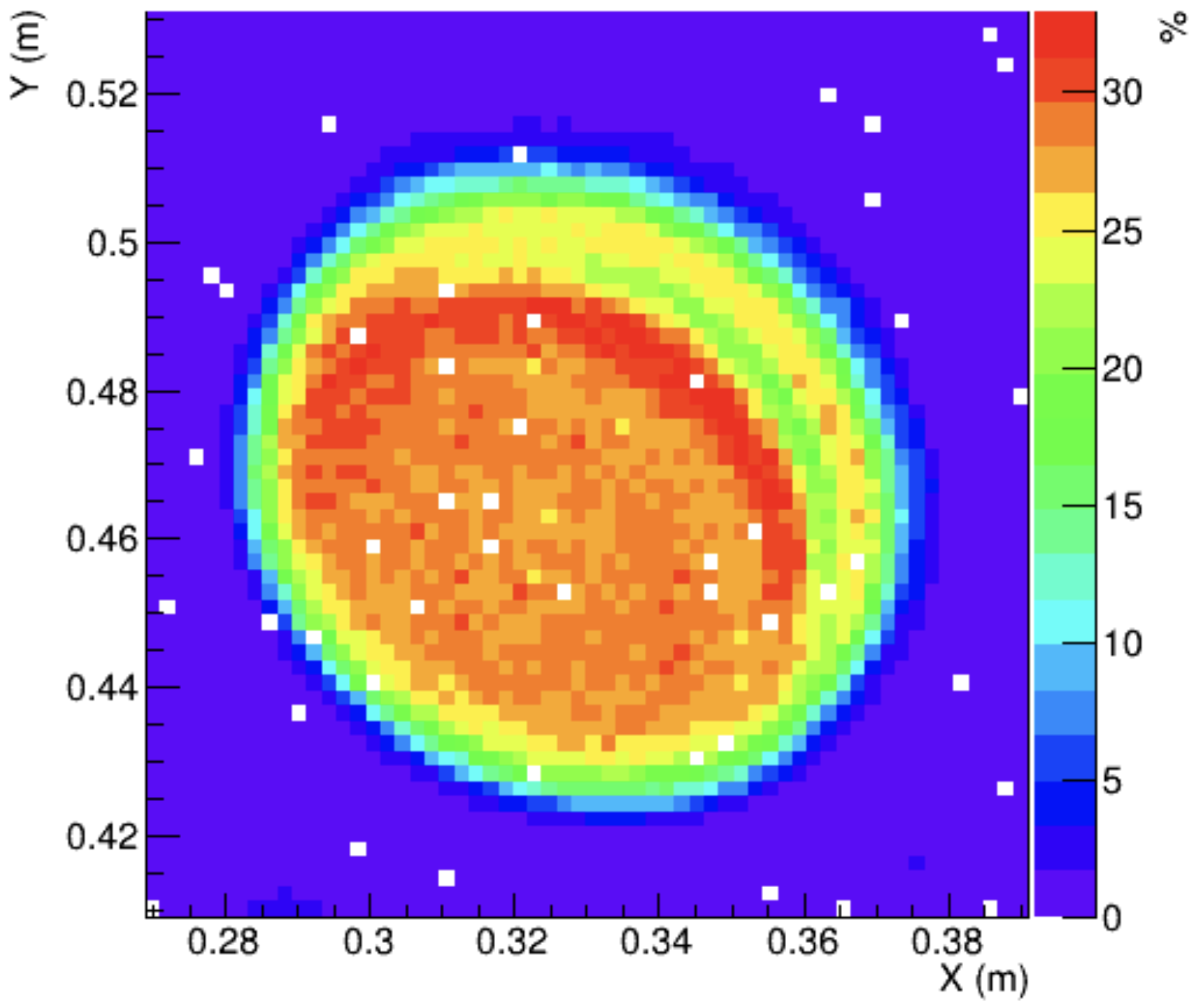}
        \caption*{(a) PMT with a reflector}
    \end{minipage}
    \hfill
    \begin{minipage}{0.45\textwidth}
        \centering
        \includegraphics[scale=0.188]{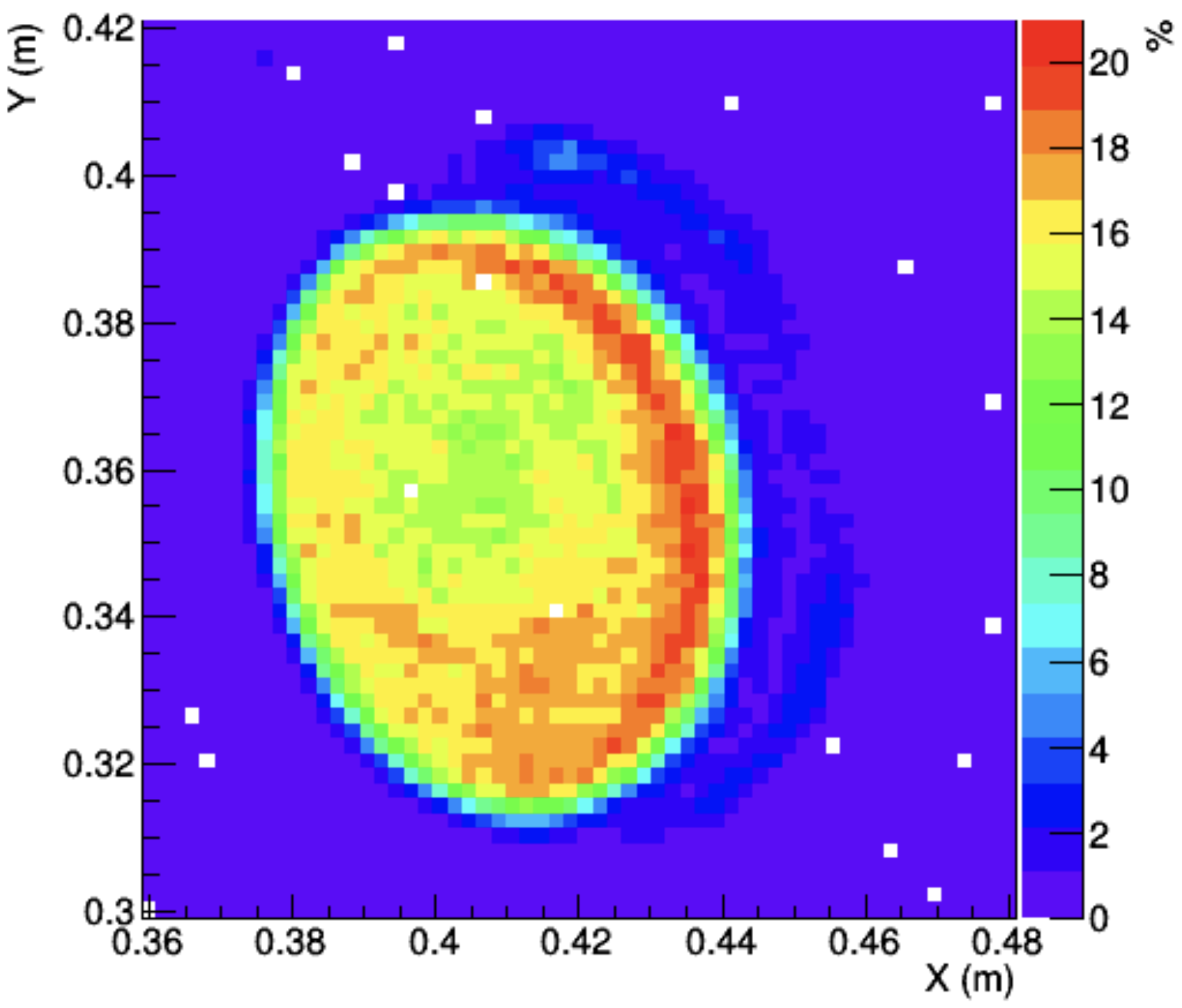}
        \caption*{(b) PMT without a reflector}
    \end{minipage}
    \caption{The plots show the efficiency measurement for the PMT in the outermost row of the module: (a) with a reflector and (b) without a reflector.}
    \label{fig:thirdRaw_PMT_Reflector_Efficiency_Comparison}
\end{figure}

A scan of the outermost-row PMTs in one of the modules was conducted too. Figure~\,\ref{fig:thirdRaw_PMT_Reflector_Efficiency_Comparison}~(a) displays an efficiency plot for the PMT containing a reflector, with an effective area of 60~$\rm cm^2$. In contrast, Figure \ref{fig:thirdRaw_PMT_Reflector_Efficiency_Comparison}~(b) shows the efficiency plot for a PMT at the same relative position in a separate module without a reflector, having an effective area of 43~$\rm cm^2$. Visually, it can be observed once again that the reflector is capable of generating PMT hits across almost the entirety of the region where the PMT cup is located. The effective area reported here is relative to a specific angle of incidence. Since the gantry is moving perpendicular to the PMT, the absolute effective area would vary. Therefore, this measurement is not an absolute effective area but rather a relative measurement. Our primary interest is in the performance difference due to the presence or absence of the reflector only.


\begin{figure}
\begin{center} \includegraphics[scale=0.24]{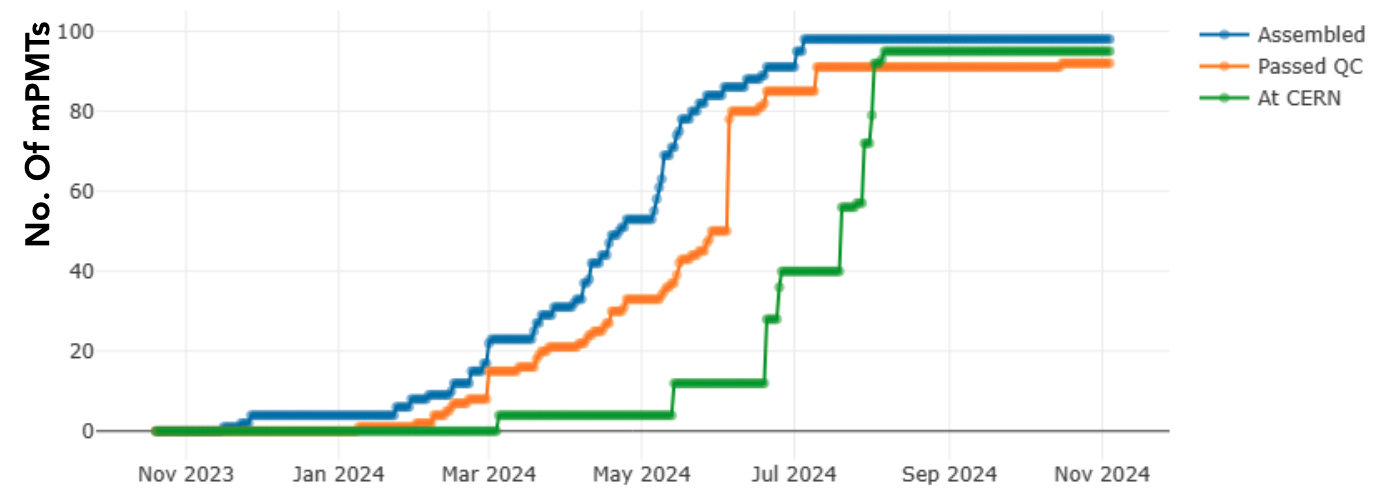} \end{center}
\caption{mPMT production summary, shows the number of mPMTs assembled, passed Quality Control (QC) and arrived at CERN for installation. }
\label{fig:mPMT_Production_Summary}
\end{figure}

\section{mPMT Production Summary}\label{s:Production_Summary}

A total of 100 IWCD-style mPMTs have been produced for the WCTE; the production was split equally between the Warsaw University of Technology and TRIUMF. The production summary, including the status of each mPMT, is shown in Figure~\,\ref{fig:mPMT_Production_Summary}. Out of these, 36 are ex-situ style and 64 are in-situ style mPMTs. During production, we encountered several issues, both mechanical and with electronics. Below, we list the mechanical problems only (electronics problems will be addressed in a separate paper):

\begin{itemize}
\item We discovered that the inner diameter of $\sim$20 cylinders was not machined correctly, causing the cylinders to rub on the PMT gel during ex-situ mPMT assembly. The inner diameter was roughly 5~$\rm mm$ less than the nominal 451.2~$\rm mm$. We performed additional machining of the inner surface of the cylinder after realizing the problem.\\

\item We have conducted a long-term test of an mPMT in a chilled water tank. The mPMT was submerged in the water for 15 months.  We observed a steady increase in humidity inside mPMT over 15 whole months, corresponding to approximately a 2\% increase in relative humidity per month. This increase is probably caused by diffusion of water, not by leakage into the mPMT vessel. This increase translates to approximately 15-30~$\rm mg$ of water per month, which would amount to 1.8-3.6~$\rm g$ over 10 years. Based on tests with other mPMTs in the same tank, we have some indication that the diffusion primarily occurs through the PVC cylinder, though this is quite uncertain. Consequently, we have begun adding desiccant packets to the assembled mPMTs. These packets are sized to absorb up to 9~$\rm g$ of water, which should be sufficient for WCTE and 20 years of IWCD operation.
\end{itemize}

\begin{figure}[!h]
\centering
\includegraphics[scale=0.0425]{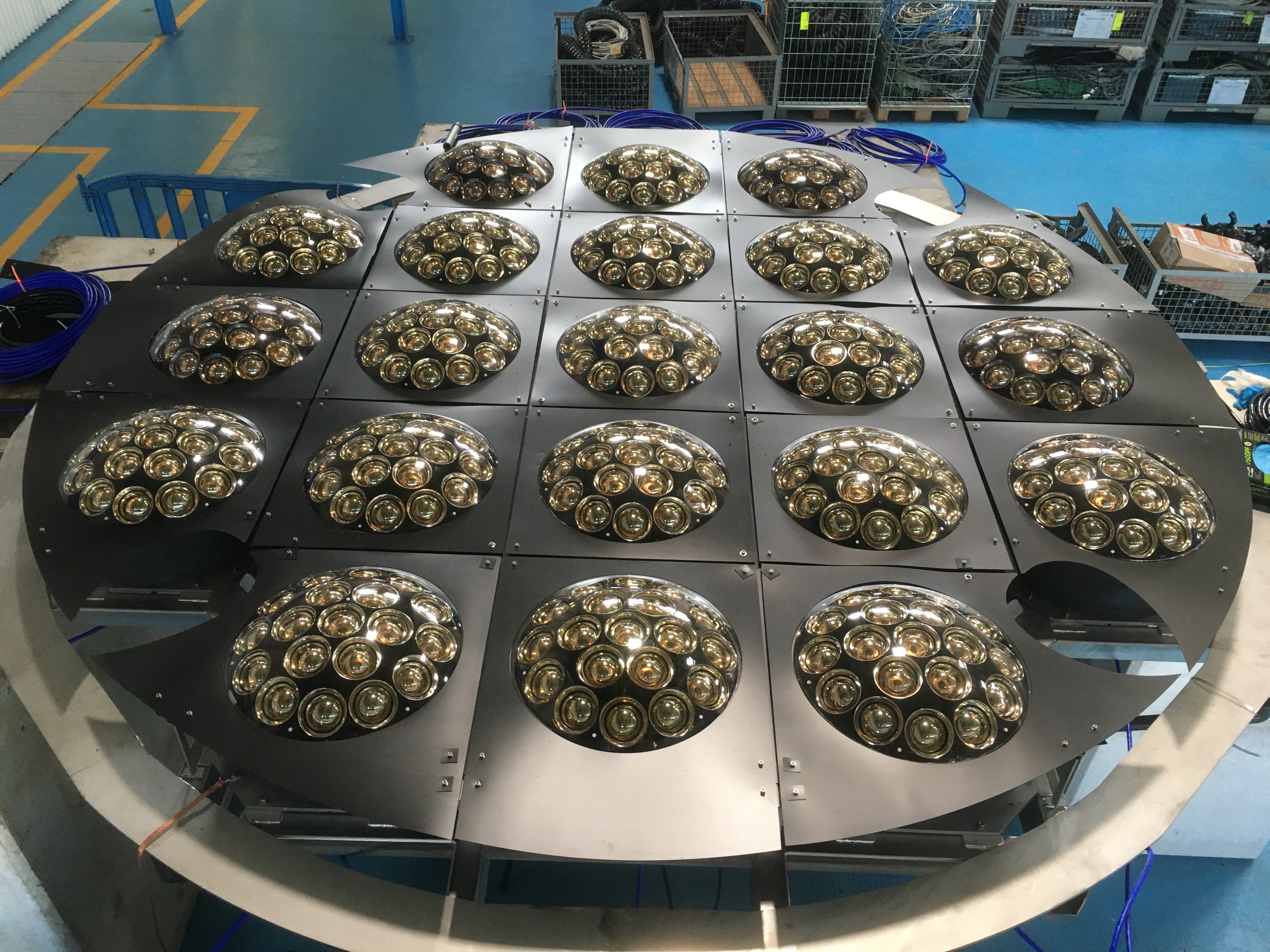}
\includegraphics[scale=0.0425]{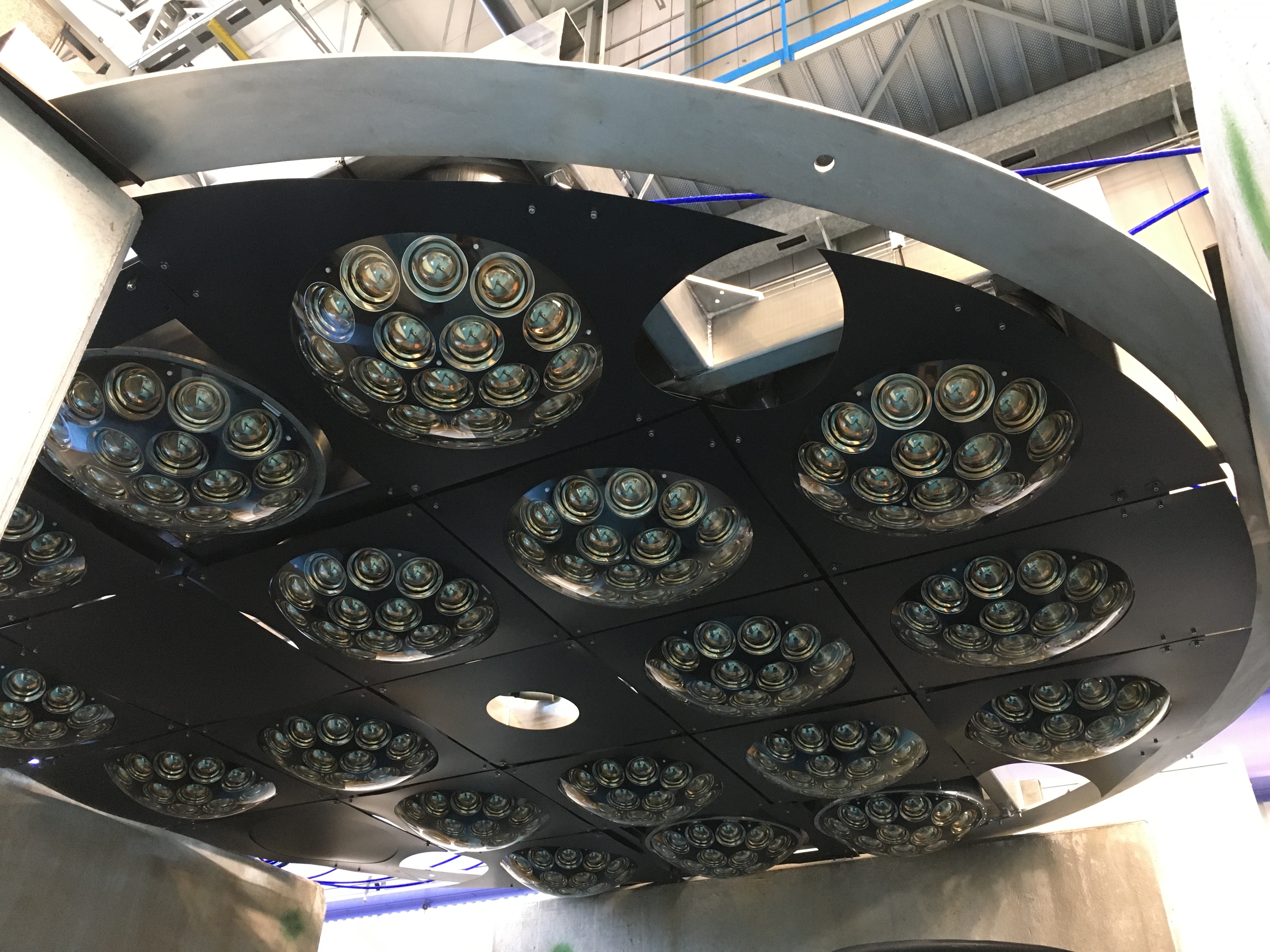}
\includegraphics[scale=0.15]{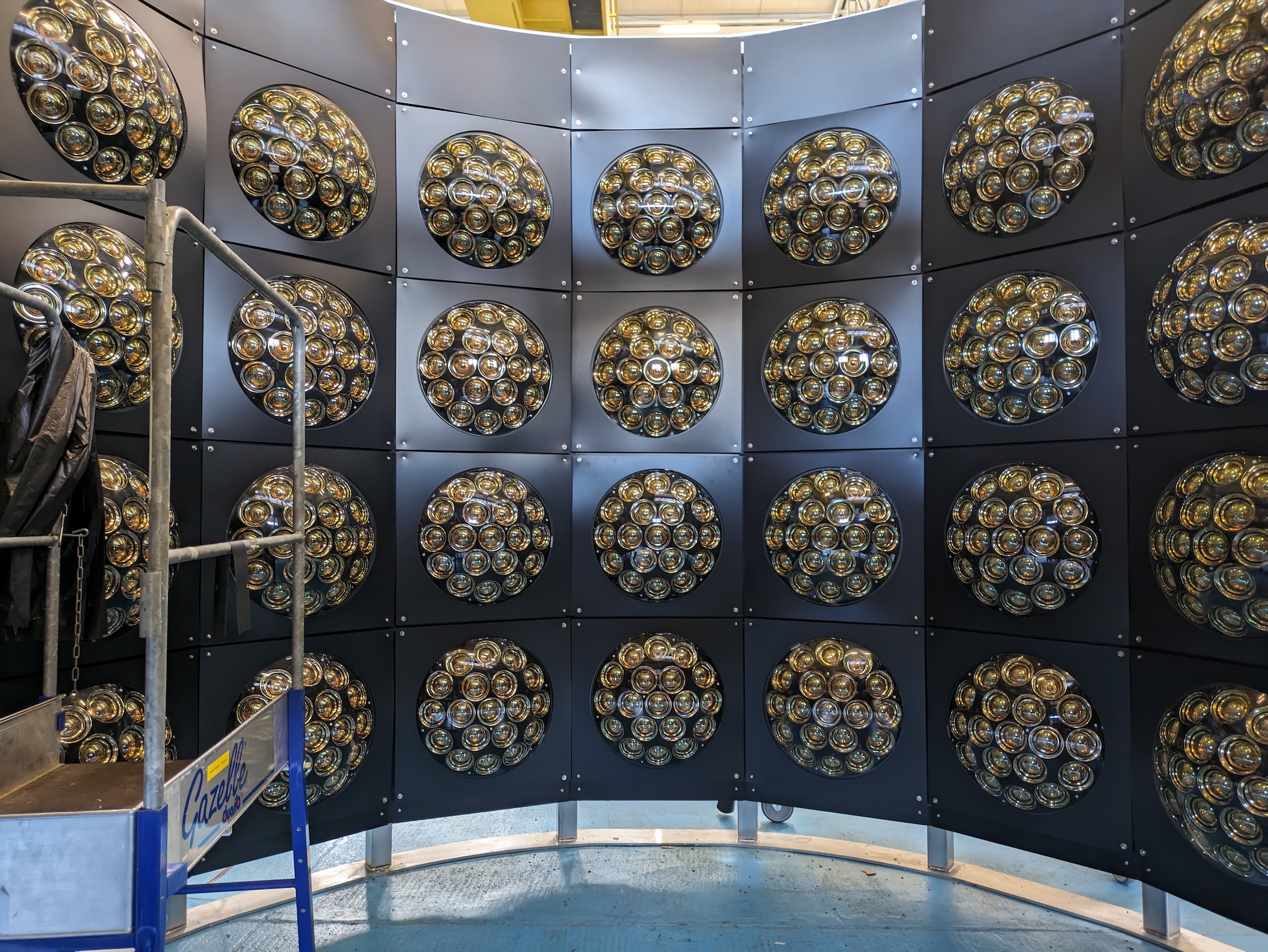}
\caption{The mPMTs installed across the WCTE support structure components: Bottom Endcap (BEC), Top Endcap (TEC), and Barrel at CERN.}
\label{fig:WCTE_Structure}
\end{figure}

Once the mPMTs pass all the necessary tests, they are shipped to CERN for installation into the WCTE structure. The installation of the mPMTs into the support structure occurs in three parts: the top endcap (TEC), the bottom endcap (BEC), and the barrel. Each endcap accommodates 21 mPMTs, and the barrel can house 64 mPMTs. However, not all slots are filled due to the limited number of mPMTs produced and the installation of other instruments (beam pipe and calibration system). Figure~\,\ref{fig:WCTE_Structure} shows all three parts separately, with the mPMTs installed in each section.

\section{Conclusions}\label{s:Conclusions}
The IWCD will serve as a near detector for the Hyper-K experiment. The mPMT photosensor has been developed for use in the IWCD due to its better timing and spatial resolution compared to the Super-K or Hyper-K 20-inch PMTs. To validate the performance of the mPMTs and their event reconstruction capabilities, a 40-ton WCTE was constructed and successfully operated at CERN from late 2024 through mid-2025. The detector is currently undergoing decommissioning. Two different assembly strategies, namely ex-situ and in-situ, have been developed to manufacture these mPMTs and establish optical contact between the PMTs and the acrylic dome cover of the mPMT vessel.  A novel assembly procedure, as well as the mechanical components involved in the assembly of the mPMT photosensors, has been discussed in this paper. Approximately 400 mPMTs will be manufactured specifically for the IWCD experiment, including the 100 mPMTs production described in this paper.  Given that the in-situ style mPMT is faster to assemble and requires less manpower, this strategy brings advantages for future IWCD mPMT production. The paper also presents the first results from a fully assembled mPMT, tested with a laser connected to a gantry system inside a dark box. A summed efficiency plot that shows the relative efficiency differences of each PMT within the module is presented. This paper also examines the incorporation of a metallic reflector around the PMT and its effect on the collection efficiency. Our observations indicate that this addition can increase the PMT's photosensitive area by up to approximately 30\%. In conclusion, a novel assembly procedure has been developed to build mPMT photosensors for the WCTE. This procedure has been refined through the production of 100 mPMTs and will be used to build approximately 300 more mPMTs for the IWCD experiment.

\section{Acknowledgements}
This project received funding from the Ministry of Science and Higher Education, Republic of Poland, "International co-financed projects", grant number 5316/H2020/2022/2023/2; This project has received funding from the European Union’s Horizon 2020 research and innovation programme under the Marie Skłodowska-Curie grant agreement No 872549.  This project received funding from the Ministry of Science and Higher Education, Republic of Poland, "Support for Polish research teams participating in international research infrastructure projects", grant number 2022/WK/15.  This research was supported by the Ministry of Education Youth and Sports of the Czech Republic (FORTE CZ.02.01.01/00/22\_008/0004632) and by Charles University (Grant PRIMUS 23/SCI/025 and Research Center UNCE/24/SCI/016).  We acknowledge the support of the Natural Sciences and Engineering Research Council of Canada (NSERC).

%
%


\begin{thebibliography}{6}
%

\bibitem {HK1}
Hyper-Kamiokande Proto-Collaboration, Hyper-Kamiokande Design Report, arXiv:1805.04163. \url{https://doi.org/10.48550/arXiv.1805.04163}.\\

\bibitem{HK2}
Hyper-Kamiokande Proto-Collaboration, Physics potential of a long-baseline neutrino oscillation experiment using a J-PARC neutrino beam and Hyper-Kamiokande. Progress of Theoretical and Experimental Physics 2015, Volume 2015, Issue 5, May 2015, 053C02. \url{https://doi.org/10.1093/ptep/ptv061}.\\

\bibitem{HK3}
K. Abe, et al., Letter of Intent: The Hyper-Kamiokande Experiment-Detector Design and Physics Potential, arXiv:1109.3262, \url{https://doi.org/10.48550/arXiv.1109.3262}.\\

\bibitem{SK1}
Super-Kamiokande Collaboration, The Super-Kamiokande detector, NIM A 501 (2003) 418-462, DOI: \url{10.1016/S0168-9002(03)00425-X}.\\

\bibitem{SK2}
K. Abe, et al., Calibration of the Super-Kamiokande detector, NIM A 737 (2014) 253-272, DOI:\url{https://doi.org/10.1016/j.nima.2013.11.081}.\\

\bibitem{SK3}
Yasuo Takeuchi, Recent results and future prospects of Super-Kamiokande, NIM A 952 (2020) 161634, DOI:\url{https://doi.org/10.1016/j.nima.2018.11.093}.\\

\bibitem{CP-Violation1}
D. V. AHLUWALIA, Y. LIU, and I. STANCU, CP-Violation in neutrino oscillations and  L/E flatness of the e-like event ratio at Super-Kamiokande, Modern Physics Letters A 17 (2002) 13-21, DOI:\url{https://doi.org/10.1142/S0217732302006138}.\\

\bibitem{CP-Violation2}
M. Aoki, K. Hagiwara, N. Okamura, Measuring the CP-violating phase by a long base-line neutrino experiment with Hyper-Kamiokande, Physics Letter B 554 (2003) 121-132, DOI:\url{https://doi.org/10.1016/S0370-2693(02)03292-6}.\\

\bibitem{protonDecay}
Francesca Di Lodovico and on behalf of the Hyper-Kamiokande Collaboration, The Hyper-Kamiokande Experiment, J. Phys.: Conf. Ser. 888 012020 (2017).\\

\bibitem{IWCD}
J. Brar, Time Calibration of IWCD: A new water Cherenkov near detector for Hyper-Kamkiokande, FGS - Electronic Theses and Practica (2019). URL:\url{http://hdl.handle.net/1993/34471}.\\

\bibitem{FD-mPMT1}
B. QUILAIN, Multi-PMT modules for Hyper-Kamiokande. JPS Conf. Proc. 27, 011017 (2019). DOI:\url{https://doi.org/10.7566/JPSCP.27.011017}.\\

\bibitem{FD-mPMT2}
G. Ross, Hyper-Kamiokande Proto-Collaboration, A multi-PMT photodetector system for the Hyper-Kamiokande experiment, NIM A 958 (2020) 163033. DOI:\url{https://doi.org/10.1016/j.nima.2019.163033}.\\

\bibitem{FD-mPMT3}
N Deshmukh et al., Mechanical design of multi-PMTs for IWCD, J. Phys.: Conf. Ser. (2022) 2374 012134 DOI:\url{10.1088/1742-6596/2374/1/012134}.\\

\bibitem{FD-mPMT4}
B. QUILAIN et al., Multi-PMT modules for Hyper-Kamiokande, JPS Conf. Proc. 27 (2019) 011017 DOI:\url{https://doi.org/10.7566/JPSCP.27.011017}.\\

\bibitem{FD-PMT}
M. Inomoto et al., Basic studies of 3-inch PMT for multi-PMT development, J. Phys.: Conf. Ser. (2020) 1468 012162 DOI:\url{10.1088/1742-6596/1468/1/012162}.\\

\bibitem{FD-mPMT-Electronics}
Luigi Lavitola on behalf of Hyper-Kamiokande collaboration, multi-PMT electronics system for Hyper-Kamiokande, NIM A 1054  (2023) 168461 DOI:\url{https://doi.org/10.1016/j.nima.2023.168461}.\\

\bibitem{KM3Net1}
KM3NeT Collaboration, The KM3NeT multi-PMT optical module, arXiv:2203.10048. DOI:\url{https://doi.org/10.48550/arXiv.2203.10048}.\\

\bibitem{KM3Net2}
Adrian-Martinez S., et al., KM3NeT Collaboration, Deep sea tests of a prototype of the KM3NeT digital optical module. Eur. Phys. J. C, 74 (9) (2014), p. 3056. DOI:\url{
https://doi.org/10.1140/epjc/s10052-014-3056-3}.\\

\bibitem{WCTE0}
M. Gola, The Water Cherenkov Test Experiment: Investigating Particle Detection in Small-Scale Water Cherenkov Detectors, Neutrino 2024 Zenodo, Milano, Italy, DOI: \url{https://doi.org/10.5281/zenodo.13463746}\\

\bibitem{WCTE1}
M. Barbi et al., Proposal for A Water Cherenkov Test Beam Experiment for Hyper-Kamiokande and Future Large-scale Water-based Detectors, tech. rep., CERN, Geneva, Mar 2020. CERN-SPSC-2020-005 ; SPSC-P-365.\\

\bibitem{WCTE2}
The WCTE Collaboration, WATER CHERENKOV TEST EXPERIMENT STATUS REPORT. CERN-SPSC-2023-016 / SPSC-SR-329. URL:\url{https://cds.cern.ch/record/2857041/files/SPSC-SR-329.pdf}.\\

\bibitem{IWCD-mPMT}
M. Gola, mPMT photosensors development for the Water Cherenkov Test Experiment (WCTE), 2023 IEEE Nuclear Science Symposium, Medical Imaging Conference and International Symposium on Room-Temperature Semiconductor Detectors (NSS MIC RTSD), Vancouver, BC, Canada, 2023, pp. 1-1, DOI: \url{10.1109/NSSMICRTSD49126.2023.10338476}.\\

\bibitem{SK-Gd}
Super-Kamiokande Collaboration, A. Goldsack for the collaboration, The New Phase of Super-Kamiokande: SK-Gd. PoS NuFact2021 (2022) 170. DOI:\url{10.22323/1.402.0170}.\\

\bibitem{20-inch PMT}
Y. O. Nishimura on behalf of the Hyper-Kamiokande Proto-collaboration, New 50-cm photo-detectors for Hyper-Kamiokande. Proceedings of 38th International Conference on High Energy Physics PoS 282, (ICHEP2016) 303. DOI:\url{ 10.22323/1.282.0303}.\\

\end{thebibliography}
\end{document}